\documentclass[draft]{agujournal2019}
\usepackage{aas_macros}
\usepackage{amsmath}
\usepackage{url} %this package should fix any errors with URLs in refs.
\usepackage{lineno}
\usepackage[finalnew]{trackchanges} %for better track changes. finalnew option will compile document with changes incorporated.
\usepackage{soul}
\usepackage{array,multirow}
\draftfalse

\journalname{Journal of Geophysical Research}

\begin{document}

%% ------------------------------------------------------------------------ %%
%  Title
%
% (A title should be specific, informative, and brief. Use
% abbreviations only if they are defined in the abstract. Titles that
% start with general keywords then specific terms are optimized in
% searches)
%
%% ------------------------------------------------------------------------ %%

% Example: \title{This is a test title}

\title{New Constraints on the Jovian Narrowband Radio Components from Juno/Waves Observations and 3D Geometrical Simulations}

%% ------------------------------------------------------------------------ %%
%
%  AUTHORS AND AFFILIATIONS
%
%% ------------------------------------------------------------------------ %%

% Authors are individuals who have significantly contributed to the
% research and preparation of the article. Group authors are allowed, if
% each author in the group is separately identified in an appendix.)

% List authors by first name or initial followed by last name and
% separated by commas. Use \affil{} to number affiliations, and
% \thanks{} for author notes.
% Additional author notes should be indicated with \thanks{} (for
% example, for current addresses).

% Example: \authors{A. B. Author\affil{1}\thanks{Current address, Antartica}, B. C. Author\affil{2,3}, and D. E.
% Author\affil{3,4}\thanks{Also funded by Monsanto.}}

\authors{A. Boudouma\affil{1, 2}, P. Zarka\affil{2}, C.K. Louis\affil{2}, M. Imai\affil{1}, C. Briand\affil{2}}

% \affiliation{1}{First Affiliation}
% \affiliation{2}{Second Affiliation}
% \affiliation{3}{Third Affiliation}
% \affiliation{4}{Fourth Affiliation}
\affiliation{1}{Department of Space Physics, Institute of Atmospheric Physics, Czech Academy of Science, Prague, Czechia}
\affiliation{2}{LIRA, Observatoire de Paris, CNRS, PSL, Sorbonne Université, Université Paris Cité, Meudon, France}
%(repeat as many times as is necessary)

%% Corresponding Author:
% Corresponding author mailing address and e-mail address:

% (include name and email addresses of the corresponding author.  More
% than one corresponding author is allowed in this LaTeX file and for
% publication; but only one corresponding author is allowed in our
% editorial system.)

% Example: \correspondingauthor{First and Last Name}{email@address.edu}

\correspondingauthor{Adam Boudouma}{boudouma@ufa.cas.cz}

%% Keypoints, final entry on title page.

%  List up to three key points (at least one is required)
%  Key Points summarize the main points and conclusions of the article
%  Each must be 140 characters or fewer with no special characters or punctuation and must be complete sentences

% Example:
% \begin{keypoints}
% \item	List up to three key points (at least one is required)
% \item	Key Points summarize the main points and conclusions of the article
% \item	Each must be 140 characters or fewer with no special characters or punctuation and must be complete sentences
% \end{keypoints}

\begin{keypoints}
\item We applied the 3D geometrical model LsPRESSO to study the generation mechanisms of the jovian narrowband components.
\item nKOM is more consistent with $\omega_{pe}$ and/or $2 \omega_{pe}$ in X-mode at low-latitude, and $\omega_{pe}$ in O-mode at high-latitude.
\item nLF is consistent with $\omega_{pe}$ and $2\omega_{pe}$ in O-mode at all latitudes, but also at $2\omega_{pe}$ in X-mode at low-latitudes.
\end{keypoints}

%% ------------------------------------------------------------------------ %%
%
%  ABSTRACT and PLAIN LANGUAGE SUMMARY
%
% A good Abstract will begin with a short description of the problem
% being addressed, briefly describe the new data or analyses, then
% briefly states the main conclusion(s) and how they are supported and
% uncertainties.

% The Plain Language Summary should be written for a broad audience,
% including journalists and the science-interested public, that will not have 
% a background in your field.
%
% A Plain Language Summary is required in GRL, JGR: Planets, JGR: Biogeosciences,
% JGR: Oceans, G-Cubed, Reviews of Geophysics, and JAMES.
% see http://sharingscience.agu.org/creating-plain-language-summary/)
%
%% ------------------------------------------------------------------------ %%

%% \begin{abstract} starts the second page

\begin{abstract}

Measurements of Waves instrument onboard the Juno spacecraft suggest that narrowband kilometric radiation (nKOM; 20-141 kHz) and narrowband low-frequency radiation (nLF; 5-70 kHz) are generated within the plasma near the Io plasma torus (IPT) in low-latitude regions. While these emissions are thought to result from the conversion of the natural modes of the plasma into escaping radio waves, at either the fundamental or the first harmonic of the plasma frequency, there is no consensus on the specific mechanism involved. Using the electron density and the magnetic field data from the Jovian Auroral Distribution Experiment (JADE) and the FluxGate Magnetometer (FGM), we determine the range of frequencies accessible to different wave modes during Juno's crossing of the plasma disk environment. We classify the observed nKOM and nLF according to their propagation modes: trapped (Z-mode or Whistler), escaping (X-mode or O-mode), and undetermined (either trapped or escaping).
We apply the 3D numerical modeling method that was developed in \citeA{Boudouma2024} to the escaping and undetermined nKOM and nLF observations, deriving macroscopic constraints on the generation mechanisms, wave modes, characteristic frequencies, beaming and source locations. Our results support the interpretation in which high-latitude nKOM is consistent with O-mode, while low-latitude is rather X-mode. Both nKOM and nLF appear to be generated near the fundamental of the plasma frequency, but only nLF shows compatibility with emission near the first harmonic, suggesting the possible coexistence of both linear and nonlinear generation mechanisms.

\end{abstract}

\section*{Plain Language Summary}
The Jupiter narrowband kilometric radiation (nKOM) and narrowband low-frequency radiation (nLF), observed by Waves instrument onboard the Juno spacecraft, are believed to be generated in the plasma disk near a region called the Io plasma torus. These radio signals are thought to be generated from the conversion of the plasma energy into electromagnetic waves that can propagate into the vaccum, but the exact process behind this transformation remains uncertain. To better understand how these emissions form, we used measurements of the local plasma and magnetic environment taken by Juno’s JADE and FGM instruments. This helped us identify which types of radio waves could exist in the observed conditions. We then applied a 3D numerical model to analyze the nKOM and nLF emissions. Our findings suggest that nKOM waves seen at high-latitudes match a type called O-mode, while those observed from low-latitudes are more likely X-mode. Both emissions seem to be generated near a key plasma frequency, but only nLF also fits with emission at double that frequency. This points to the possibility of different generation processes occurring together.

%% ------------------------------------------------------------------------ %%
%
%  TEXT
%
%% ------------------------------------------------------------------------ %%

%%% Suggested section heads:
% \section{Introduction}
%
% The main text should start with an introduction. Except for short
% manuscripts (such as comments and replies), the text should be divided
% into sections, each with its own heading.

% Headings should be sentence fragments and do not begin with a
% lowercase letter or number. Examples of good headings are:

% \section{Materials and Methods}
% Here is text on Materials and Methods.
%
% \subsection{A descriptive heading about methods}
% More about Methods.
%
% \section{Data} (Or section title might be a descriptive heading about data)
%
% \section{Results} (Or section title might be a descriptive heading about the
% results)
%
% \section{Conclusions}

\section{Introduction}

Non-thermal jovian narrowband radiations below 10~MHz were discovered by Voyager Planetary Radio Astronomy (PRA) and Plasma Wave System (PWS) experiments \cite{Warwick1977, Warwick1979a, Warwick1979b, Kaiser&Desch1980, Gurnett1983, Daigne&Leblanc1986}.
%%%%%%%%%%%%%%%%%%%%%%%%%%%%%%%%%%%%%%%%%%%%%%%%%%%%%%%%%%%%%%%%%%%%%%%%%%%%%
%_nKOM:
%   A) morphologie: 
%       _blobs lisses, periodiques ~10h (>periode corotation)
%       _gamme de fréquence 60 kHz - 160 kHz
%       _Dominant O-mode et avec X-mode (Kaiser 1980, Carr 1983, Daigne 1986, Reiner 1993)
%   B) generation:
%        _Dans IPT, basses latitudes (Reiner et al. 1993, Imai et al. 2017)
They mostly appear as narrowband kilometric radiation (nKOM) in the range 60--200~kHz \cite{Warwick1979b, Daigne&Leblanc1986, Stone1992, Reiner1993, Louarn1998, Louarn2000, Louarn2001, Zarka2004, Imai2017, Louis2021}. nKOM polarization is predominantly left-handed ordinary mode (O-mode) \cite{Daigne&Leblanc1986, Carr1983, Stone1992}, although occurrences in right-handed extraordinary mode (X-mode) have also been reported \cite{Reiner1993}. The sources of nKOM are distributed at low centrifugal latitudes within the plasma environment nearby the Io Plasma Torus (IPT) \cite{Reiner1993, Imai2017} and their modulation periods are between 3\% and 5\% longer than Jupiter's rotation period, indicating sub-corotation of the radio sources within the IPT \cite{Kaiser&Desch1980, Stone1992, Reiner1993, Louarn1998, Louarn2000, Louarn2001}. Their activation is consistent with intermittent magnetospheric perturbations that induce centrifugal plasma ejection of plasma in the plasma disk \cite{Louarn1998, Louarn2000, Louarn2001, Louarn2014}.
%%%%%%%%%%%%%%%%%%%%%%%%%%%%%%%%%%%%%%%%%%%%%%%%%%%%%%%%%%%%%%%%%%%%%%%%%%%%%

%%%%%%%%%%%%%%%%%%%%%%%%%%%%%%%%%%%%%%%%%%%%%%%%%%%%%%%%%%%%%%%%%%%%%%%%%%%%%
%_nLF:
%   A) morphologie:
%       _pareil que nKOM mais basses-fréquences 5 kHz - 70 kHz
%   B) generation:
%       _Anciennes observations => radio sources proposées à la magnetopause (Kurth 1992)
%       _Observation Juno/Waves:
%           i) Juno/Waves direction finding => suggère des sources dans l'IPT (Imai et al. 2017)
Narrowband low-frequency radiation (nLF), also referred to as escaping continuum radiation (ECR) or very low-frequency radiation (VLF), has been observed in the 5--60~kHz range \cite{Gurnett1983, Stone1992, Kaiser1992, Imai2017, Louis2021, Menietti2023}. Prior to the Juno mission, only a few nLF events had been observed \cite{Gurnett1983, Stone1992, Kaiser1992}, as its occurrence probability is relatively low at low-latitudes compared to nKOM \cite{Louis2021}. Consequently, nLF polarization remains unknown as Juno/Waves does not provide polarization measurements. The source locations of nLF are yet to be confirmed, but limited direction finding measurements suggest that they are distributed at 5$^{\circ}$--10$^{\circ}$ above the centrifugal equator near the outer edge of the plasma disk \cite{Imai2017}.
%%%%%%%%%%%%%%%%%%%%%%%%%%%%%%%%%%%%%%%%%%%%%%%%%%%%%%%%%%%%%%%%%%%%%%%%%%%%%

%%%%%%%%%%%%%%%%%%%%%%%%%%%%%%%%%%%%%%%%%%%%%%%%%%%%%%%%%%%%%%%%%%%%%%%%%%%%%
%_nKOM & nLF => mécanisme génération = mécanismes de conversion de modes
Both nKOM and nLF are ``plasma emissions'' and their generation is attributed to the conversion of the natural modes of the plasma (i.e, electrostatic or trapped electromagnetic waves excited by plasma instabilities) into escaping radio waves at the fundamental or harmonic of the plasma frequency \cite{Jones1987, Fung&Papadopoulos1987, Reiner1993, Imai2017}. However, no consensus exists on the nature of the involved mode conversion mechanisms (e.g., linear conversion, waves coupling, wave decay) or on the \change{proporties}{properties} of the resulting radio waves (e.g., frequency, beaming, polarization).
%%%%%%%%%%%%%%%%%%%%%%%%%%%%%%%%%%%%%%%%%%%%%%%%%%%%%%%%%%%%%%%%%%%%%%%%%%%%%

%%%%%%%%%%%%%%%%%%%%%%%%%%%%%%%%%%%%%%%%%%%%%%%%%%%%%%%%%%%%%%%%%%%%%%%%%%%%%
%_Boudouma et al. 2024 => methode pour contraindre les emissions plasma 
In \citeA{Boudouma2024}, hereafter referred to as B24, we proposed a numerical method to constrain plasma emissions characteristics based on their large-scale latitudinal beaming. To do so, we developed a 3D geometrical model of the ``Large-scale Plasma Radio Emissions Simulation of Spacecraft Observation'' (LsPRESSO).
%%%%%%%%%%%%%%%%%%%%%%%%%%%%%%%%%%%%%%%%%%%%%%%%%%%%%%%%%%%%%%%%%%%%%%%%%%%%%
%_Précédement appliqué à la caractérisation du nKOM
%_Résultats: nKOM à fpe avec r || -grad(ne), 2 modes:
%   A) O-mode: hautes latitudes, basses-fréquences à bande étroite 30 kHz - 50 kHz
%   B) X-mode: basses latitudes, hautes-fréqunces à bande large >20 kHz
We used LsPRESSO to constrain the characteristic frequency, large-scale beaming and cutoff of nKOM based on the Juno/Waves observations \cite{Louis2021}. We showed that the distribution of nKOM occurrence probability versus latitude and frequency is only consistent with radio waves near the fundamental of the local plasma frequency, with its beaming directed anti-parallel to the local density gradient. Furthermore, we suggested that nKOM observed by Juno/Waves corresponds to O-mode at high-latitude and X-mode at low-latitude.
%%%%%%%%%%%%%%%%%%%%%%%%%%%%%%%%%%%%%%%%%%%%%%%%%%%%%%%%%%%%%%%%%%%%%%%%%%%%%

%%%%%%%%%%%%%%%%%%%%%%%%%%%%%%%%%%%%%%%%%%%%%%%%%%%%%%%%%%%%%%%%%%%%%%%%%%%%%
%_Appliquer la méthode à la caractérisation du nLF
Here, we propose to extend this study to the characterization of the nLF. Additionally, we use measurements from the Juno Jovian Aurora Distribution Experiment (JADE) \cite{McComas2017} and the Flux Gate Magnetometer (FGM) \cite{Connerney2017} to investigate the characteristics of the medium when Juno is crossing the plasma environment near the IPT and derive further observational constraints on the propagation modes of nKOM and nLF, before comparing their distributions with LsPRESSO results.

%%%%%%%%%%%%%%%%%%%%%%%%%%%%%%%%%%%%%%%%%%%%%%%%%%%%%%%%%%%%%%%%%%%%%%%%%%%%%
%_Sommaire: 
%   Section +1: Mécanismes de conversion de modes, LMCW et mécanismes non-linéaires, Charactéristiques du plasma
%   Section +2: Missio Juno, Observations
%   Section +3: Construction des distributions en latitudes et fréquences
%   Section +4: Simulations 0) Rappel méthode, 1) nKOM & nLF (total), 2) nKOM & nLF (indeterminés et non-piégées)
%%%%%%%%%%%%%%%%%%%%%%%%%%%%%%%%%%%%%%%%%%%%%%%%%%%%%%%%%%%%%%%%%%%%%%%%%%%%%
Section \ref{sec:mode_conversion} presents the key aspects of the mode conversion mechanisms involved in the generation of radio waves within the plasma disk. Section \ref{sec:juno_data} describes the Juno/Waves, JADE and FGM data used in this study and the observed distributions that we aim to reproduce in our modeling. Section \ref{sec:LsPRESSO} reviews the B24 method and presents the results from the LsPRESSO modeling. Section \ref{sec:comparison} compares the results obtained with the Juno/Waves observations on the first perijove. Section \ref{sec:discussion} summarizes the results obtained and discusses the relevance of the potential mode conversion mechanisms in the generation of nKOM and nLF.

\section{Generation of X- and O-modes \change{in the Io's}{in Io's} Plasma Torus}\label{sec:mode_conversion}

Voyager, Ulysses and Juno observations suggest that the generation of nKOM and nLF takes place near the IPT. The electron plasma frequency $\omega_{pe}$ and the electron cyclotron frequency $\omega_{ce}$ are: 

\begin{equation}
    \omega_{pe}~=~\sqrt{\frac{n_e e^2}{m_e \epsilon_0}}
\end{equation}

\begin{equation}
     \omega_{ce}~=~\frac{eB}{m_e}
\end{equation}

where $n_e$, $e$ and $m_{e}$ are the electron density, charge and mass, and $\epsilon_0$ is the vacuum permittivity and $B$ is the magnetic amplitude. In the IPT $\omega_{pe} \ge \omega_{ce}$ or $\omega_{pe} \sim \omega_{ce}$, favoring mode conversion mechanisms over the electron cyclotron maser instability \cite{Perraut1998, Treumann2006}.
%%%%%%%%%%%%%%%%%%%%%%%%%%%%%%%%%%%%%%%%%%%%%%%%%%%%%%%%%%%%%%%%%%%%%%%%%%%%%
%_Mécanismes de conversion de mode: 
%   A) linéaire => conversion de l'énergie d'un mode vers un autre
%   B) non-linéaire => couplage d'onde(s) ou diffraction
Several mode conversion mechanisms, both linear and nonlinear, are presently debated to explain the generation of escaping radio waves (X- and O-modes) in the IPT.
%%%%%%%%%%%%%%%%%%%%%%%%%%%%%%%%%%%%%%%%%%%%%%%%%%%%%%%%%%%%%%%%%%%%%%%%%%%%%

%%%%%%%%%%%%%%%%%%%%%%%%%%%%%%%%%%%%%%%%%%%%%%%%%%%%%%%%%%%%%%%%%%%%%%%%%%%%%
%_Mécanismes de conversion de mode linéaire => génération à fpe suivant conversion linéaire de l'énergie d'un mode vers un/plusieurs autre(s)
%_Principe:
Linear mode conversion (LMC) mechanisms involve the linear transfer of energy from one mode to another. They can explain the generation of escaping radio waves at the fundamental of $\omega_{pe}$. 
%%%%%%%%%%%%%%%%%%%%%%%%%%%%%%%%%%%%%%%%%%%%%%%%%%%%%%%%%%%%%%%%%%%%%%%%%%%%%
%_Cas de Jupiter: LMCW de Jones, conversion Z-mode vers O-mode à fpe dans un plasma froid non collisionnel et homogène quand alpha ~ 90°. Radio sources à l'équateur et free-space beaming dépend d'un beta = arctan(sqrt(fpe/fce)). 
Considering the Jovian case, \citeA{Jones1987} proposed the linear conversion of Z-mode into O-mode near $\omega_{pe}$ to explain the generation of planetary radio emissions including the nKOM. By assuming a cold collisionless magnetized plasma with weak $\nabla n_e$ near the conversion region, he predicts the generation of O-mode for $\nabla n_e \perp {\bf B}$. The rays are beamed in two opposite directions making angles $\beta$ and $\pi~-~\beta$ with respect to $\bf{B}$ in the plane defined by ($\nabla n_e$, $\bf{B}$), with $\beta = \arctan (\sqrt{\omega_{pe}/\omega_{ce}})$. Thus, the radio sources are expected to lie near the equatorial plane where the condition $\nabla n_e \perp {\bf B}$ is more likely to be fulfilled. 
%%%%%%%%%%%%%%%%%%%%%%%%%%%%%%%%%%%%%%%%%%%%%%%%%%%%%%%%%%%%%%%%%%%%%%%%%%%%%
%_Theorie Jones mise en difficulté: 
%   1) efficacité de la conversion trop faible 
%   2) précédente étude montre Jones incompatible avec distribution du nKOM
However, the Jones LMC theory has been challenged as its conversion efficiency and beaming are incompatible with the observations of the terrestrial continuum radiation (TCR) \cite{Melrose1981, Barbosa1982, Ronnmark1989, Boardsen2023}. Furthermore, the modeling presented in B24 also suggests that this theory is incompatible with the Juno/Waves nKOM observations.
%%%%%%%%%%%%%%%%%%%%%%%%%%%%%%%%%%%%%%%%%%%%%%%%%%%%%%%%%%%%%%%%%%%%%%%%%%%%%
%_Conversion linéaire => gradient de densité: meilleure éfficacité
Nevertheless, LMC mechanisms present higher conversion efficiencies when considering the plasma inhomogeneous in the conversion region \cite{AshourAbdalla1984, Budden&Jones1987}. 
%%%%%%%%%%%%%%%%%%%%%%%%%%%%%%%%%%%%%%%%%%%%%%%%%%%%%%%%%%%%%%%%%%%%%%%%%%%%%
%_Efficacité du mécanisme et polarization on été largement étudié au travers de simulations (Schleyer 2014 et toutes les refs dedans) :
%   1) angle entre grad(ne) et B => Emax quand perp, Emin quand par
%   2) température des électrons => + chaud mieux c'est
%   3) force de grad(ne)) et de intensité de B par rapport à la longueur d'onde => compliqué
%_Polarization dépend de: force de grad(ne)) et de intensité de B par rapport à la longueur d'onde
%_Asymptotiquement:
%   1) Plasma très mangétisé et/ou grad(ne) très faible => seulement O-mode
%   2) Plasma faiblement magnétisé et/ou grad(ne) très fort  => X- et O-modes
Numerical studies have explored the linear conversion of Langmuir waves into escaping electromagnetic waves at $\omega_{pe}$ in a warm collisionless inhomogeneous magnetized plasma \cite{Cairns&Willes2005, Kim2007, Kim2008, Kim2013, Schleyer2013, Schleyer2014}. They have shown that the conversion is more efficient for plasmas with high electron temperatures $T_{e}$ and for specific angles between $\nabla n_e$ and ${\bf B}$, depending on generated modes and density variation length scale \cite{Schleyer2013, Schleyer2014}. They predict the generation of both X- and O-modes in weakly magnetized plasma, and the generation of O-mode only in magnetized plasmas \cite{Kim2007, Kim2008, Kim2013}.
%%%%%%%%%%%%%%%%%%%%%%%%%%%%%%%%%%%%%%%%%%%%%%%%%%%%%%%%%%%%%%%%%%%%%%%%%%%%%

%%%%%%%%%%%%%%%%%%%%%%%%%%%%%%%%%%%%%%%%%%%%%%%%%%%%%%%%%%%%%%%%%%%%%%%%%%%%%
%_Mécanismes de conversion de mode non-linéaire => génération à fpe ou harmoniques
Nonlinear mode conversion (NLMC) mechanisms involve the nonlinear coupling between multiple modes. They can explain the generation of escaping radio waves at both the fundamental and the harmonics of $\omega_{pe}$.
%%%%%%%%%%%%%%%%%%%%%%%%%%%%%%%%%%%%%%%%%%%%%%%%%%%%%%%%%%%%%%%%%%%%%%%%%%%%%
%_Cas de Jupiter: Fung & Papadopoulos, coalescence d'onde L à fuh vers X- et O-mode à 2fuh dans un plasma froid non collisionnel et homogène. Beaming dans les directions perp à B.
Considering the Jovian case, \citeA{Fung&Papadopoulos1987} proposed the nonlinear coupling between two Langmuir waves initially at the upper-hybrid frequency $\omega_{uh}=\sqrt{\omega_{pe}^2 + \omega_{ce}^2}$, into X- and O-modes at $2f_{uh}$ beamed perpendicularly to ${\bf B}$, to explain the generation of nKOM. They assumed a cold collisionless magnetized plasma with weak $\nabla n_e$. This generation scenario has also been explored in B24 and we show that it cannot explain the Juno/Waves nKOM observations.
%%%%%%%%%%%%%%%%%%%%%%%%%%%%%%%%%%%%%%%%%%%%%%%%%%%%%%%%%%%%%%%%%%%%%%%%%%%%%
%_Principe => compliqué
%_Etudié au travers de simulations dans différentes conditions:
%   _Plasma homogène => onde H (Henri et al. 2019)
%   _Plasma inhomogène non mangétisé => onde F et H
%   _Plasma inhomogène et magnétisé => nstabilité de Weibel ; 2 cas:
%       a) Plasma fpe/fce << 1 => Weibel amplifie la generation à fpe => onde F
%       b) Plasma fpe/fce ~ 1 => Weibel amortie la generation à fpe => onde H
Extensive numerical studies have been performed on the nonlinear coupling between beam-driven Langmuir waves into electromagnetic waves as they are the most favored mechanism to explain the generation of type II and III solar bursts \cite{Thurgood&Tsiklauri2015, Henri2019, Krafft&Savoini2021, Bacchini&Philippov2024, Krafft&Volokitin2024}. Unlike its linear counterpart, this conversion process can only occur if the Langmuir wave amplitude reaches a given threshold \cite{Henri2010, Briand2014} and relies on very different parameters depending on the properties of the plasma. They have shown that in a homogeneous plasma, the conversion at $2\omega_{pe}$ is more dominant than the conversion at $\omega_{pe}$ and its efficiency depends on the ratio between electron and ion temperature $T_e/T_i$ \cite{Henri2019, Bacchini&Philippov2024}. Also, in the presence of small-scale density fluctuations\change{ favors the conversion at $\omega_{pe}$ compared to $2\omega_{pe}$ as}{, the conversion at $\omega_{pe}$ is favored in comparison to $2\omega_{pe}$, since} linear conversion occurs simultaneously \cite{Krafft&Savoini2021, Krafft&Volokitin2024, Krafft2024}.

%%%%%%%%%%%%%%%%%%%%%%%%%%%%%%%%%%%%%%%%%%%%%%%%%%%%%%%%%%%%%%%%%%%%%%%%%%%%%

\section{Data}\label{sec:juno_data}

%JUNO
%Juno Mission: orbite, période, survols
%Juno/Waves (Kurth et al. 2017) antennes dipolaires + bobine magnetique
%Juno JADE: ???

%%%%%%%%%%%%%%%%%%%%%%%%%%%%%%%%%%%%%%%%%%%%%%%%%%%%%%%%%%%%%%%%%%%%%%%%%%%%%

\begin{figure}[!ht]
\centering
\noindent\includegraphics[width=\textwidth]{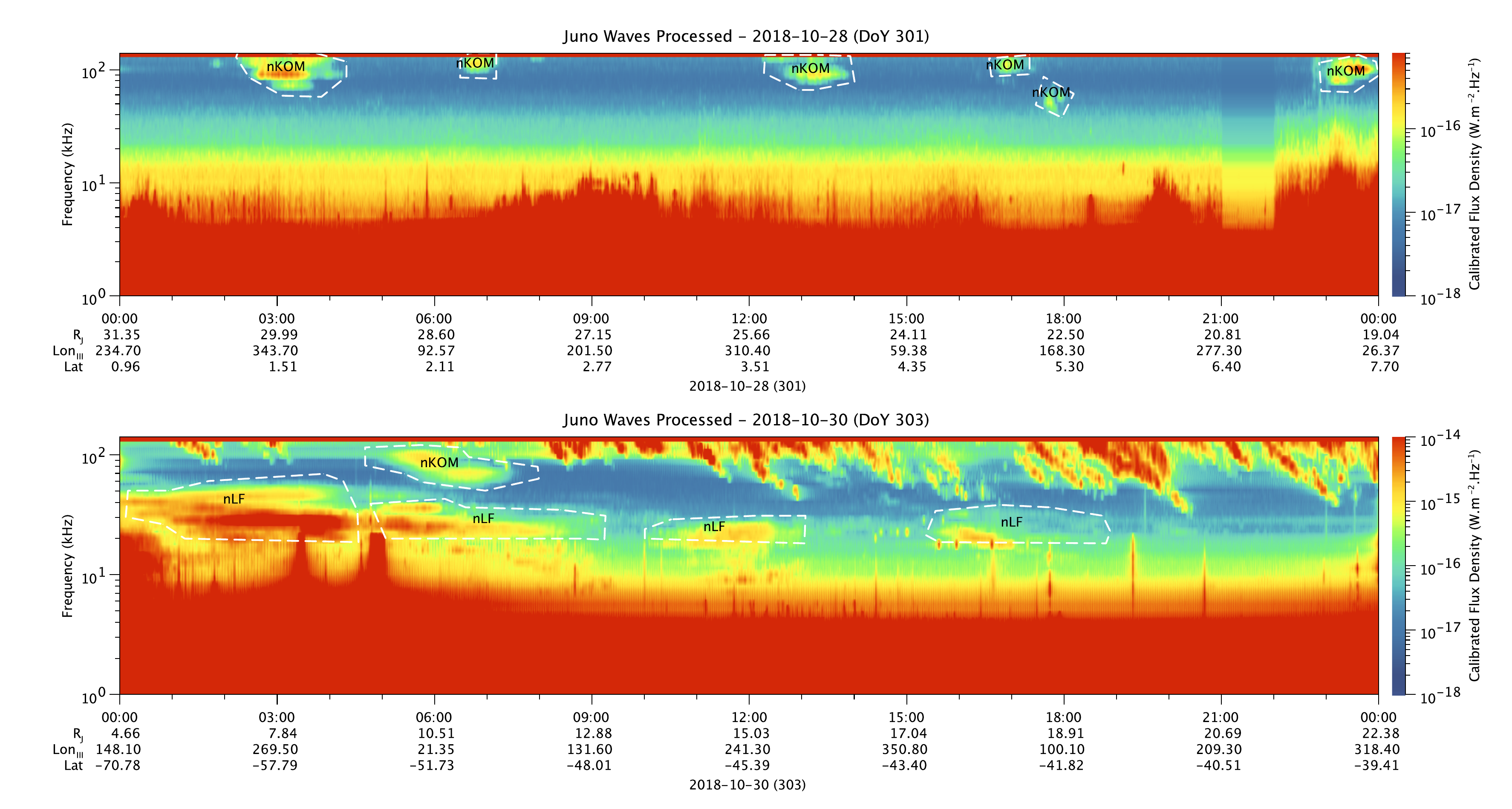}
\caption{24-hour dynamic spectra of calibrated Juno/Waves data \cite{Louis2021}  observed around the 16th perijove PJ16 (top-panel) a day before from low-latitudes (day-of-year 2018-301, i.e., 28/10/2018) and (bottom-panel) a day after from high-latitudes (day-of-year 2018-303, i.e., 30/10/2018). The white boxes correspond to the catalogued nKOM and nLF components with their labels displayed in black \cite{Louis2021_catalog}. The horizontal axis corresponds to the Juno observations time period, radial distance (R$_{\mathrm{j}}$), system III longitude (Lon$_{\mathrm{III}}$) and jovicentric latitude (Lat). The vertical axis corresponds to the Juno/Waves frequency channels from 1 kHz to 141 kHz, distributed on a logarithmic scale with a ratio $\times 1.12$ between consecutive channels. The saturated band at 140 kHz correspond to the channel 61 of the HFR-low receiver that has a poor sensitivity.}
\label{fig:spdyn_juno}
\end{figure}

%_nKOM+nLF - Juno/Waves observations: morphologie, similitudes et différences
Figure \ref{fig:spdyn_juno} displays 24-h events of Juno/Waves typical observations one day before and after a perijove (here PJ16). Multiple nKOM and nLF are observed. nKOM is recognizable by its narrowband smooth and fuzzy time-frequency morphology. nLF also displays a similar time-frequency morphology. In comparison to nKOM, nLF patches are observed to be more intense and at lower frequencies. Among the nKOM events visible on Figure \ref{fig:spdyn_juno} top panel, two are observed 10.3 h--10.5 h apart in the range 60 kHz to 140 kHz. Similarly, for the nLF events visible on Figure \ref{fig:spdyn_juno} bottom panel, two are also observed $\sim$10h apart in the range 20 kHz--30 kHz. While the periodicity of the nKOM is caused by the sub co-rotation of the nKOM radio sources in the IPT \cite{Kaiser1980, Reiner1993, Louarn1998, Louarn2000}, the cause of the periodicity of the nLF is unknown \cite{Kaiser1992}.

%%%%%%%%%%%%%%%%%%%%%%%%%%%%%%%%%%%%%%%%%%%%%%%%%%%%%%%%%%%%%%%%%%%%%%%%%%%%%
\subsection{Latitudinal Beaming of the Jovian Narrowband Radio Components}\label{sec:juno_data:lat_freq_dist}
%%%%%%%%%%%%%%%%%%%%%%%%%%%%%%%%%%%%%%%%%%%%%%%%%%%%%%%%%%%%%%%%%%%%%%%%%%%%

\begin{figure}[!ht]
\centering
\noindent\includegraphics[width=\textwidth]{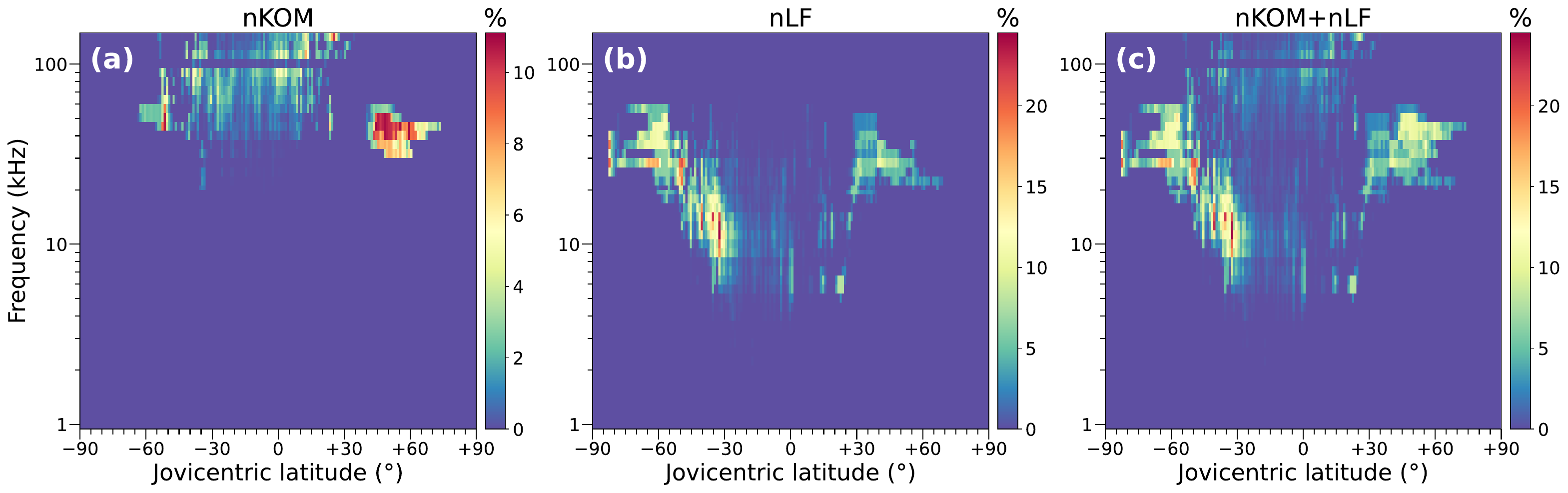}
\caption{(a) nKOM, (b) nLF and (c) nKOM+nLF occurrence probability versus frequency and latitude derived from the first 3 years of Juno/Waves low-frequency observations around Jupiter by \citeA{Louis2021}. \change{}{The color scale is adjusted in each panel and represents the occurrence probability in \%. }
The latitude correspond to the jovicentric-latitude of observation of the Juno spacecraft and the bins are 1° wide. The frequencies correspond to the Juno/Waves frequency channels from 1 kHz to 141 kHz, distributed on a logarithmic scale with a ratio $\times 1.12$ between consecutive channels. 
Panels (a) and (b) are based on the same data as Figure S4a and S4b from \citeA{Louis2021}, except that no smoothing has been applied.}
\label{fig:lat_vs_freq_dist}
\end{figure}

%%%%%%%%%%%%%%%%%%%%%%%%%%%%%%%%%%%%%%%%%%%%%%%%%%%%%%%%%%%%%%%%%%%%%%%%%%%%%
%_nKOM+nLF lat. vs. freq. distributions: nKOM+nLF complémentaires
A catalog of all the jovian radio components observed by Juno/Waves from April 9, 2016 to June 24, 2019 has been \change{dressed}{compiled} based on the emissions time-frequency morphology \cite{Louis2021_catalog, Louis2021}. About $\sim$2000 nKOM and $\sim$1200 nLF events were identified in that time period. \citeA{Louis2021} used the catalog to characterize the latitudinal variation of all the jovian radio components, including the nKOM and the nLF. Figure \ref{fig:lat_vs_freq_dist} displays the occurrence probability of (a) nKOM, (b) nLF and (c) nKOM+nLF with respect to the latitude and frequency of observation of Juno/Waves.
%%%%%%%%%%%%%%%%%%%%%%%%%%%%%%%%%%%%%%%%%%%%%%%%%%%%%%%%%%%%%%%%%%%%%%%%%%%%%
The nKOM distribution (Figure~\ref{fig:lat_vs_freq_dist}.a) shows a strong North-South asymmetry and is concentrated above $30$ kHz, with peaks in the North between $+40^{\circ}$ and $+75^{\circ}$ ($30–60$ kHz), and in the South between $-65^{\circ}$ and $-50^{\circ}$ ($40–100$ kHz). Occurrence minima appear near the equator but are only connected to the Southern peak.
%The distribution of nKOM (Figure \ref{fig:lat_vs_freq_dist}.a) shows a strong North-South asymmetry and is primarily distributed at frequencies $> 30$~kHz. It features a peak occurrence in the North at latitudes between $+40^{\circ}$ and $+75^{\circ}$ within a narrow frequency range of $30$~kHz to $60$~kHz. Another peak occurrence is observed in the South at latitudes between $-65^{\circ}$ and $-50^{\circ}$, spanning a broader frequency range from $40$~kHz to $100$~kHz. Additionally, the distribution presents occurrence minima around the equator between $-45^{\circ}$ and $+30^{\circ}$ over a wide frequency range $> 25$~kHz. However, while the equatorial occurrence minima appear connected to the Southern occurrence peak, they seem disconnected from the Northern occurrence peak.
%%%%%%%%%%%%%%%%%%%%%%%%%%%%%%%%%%%%%%%%%%%%%%%%%%%%%%%%%%%%%%%%%%%%%%%%%%%%%
The nLF distribution (Figure~\ref{fig:lat_vs_freq_dist}.b) is also asymmetric, mostly below $60$ kHz, with a broad Southern peak ($-85^{\circ}$ to $-25^{\circ}$) and a narrower Northern peak ($+25^{\circ}$ to $+70^{\circ}$). Equatorial minima are observed between $-20^{\circ}$ and $+25^{\circ}$, partially mirroring the pattern seen in nKOM.
%The distribution of nLF (Figure \ref{fig:lat_vs_freq_dist}.b) also shows a strong North-South asymmetry, but in the opposite sense compared to nKOM, and is primarily distributed at lower frequencies ranging from $5$~kHz to $60$~kHz. It shows peak occurrences in the South, spanning latitudes between $-85^{\circ}$ and $-25^{\circ}$, over a broad frequency range from $6$~kHz to $60$~kHz. Another peak occurrence is observed in the North, between latitudes $+25^{\circ}$ and $+70^{\circ}$, within a narrower frequency range of $20$~kHz to $60$~kHz. Additionally, occurrence minima are distributed around the equator, between $-20^{\circ}$ and $+25^{\circ}$, over a frequency range of $5$~kHz to $30$~kHz. Similarly to nKOM, the equatorial occurrence minima appear connected to the Southern occurrence peak but seem disconnected from the Northern occurrence peak.
%%%%%%%%%%%%%%%%%%%%%%%%%%%%%%%%%%%%%%%%%%%%%%%%%%%%%%%%%%%%%%%%%%%%%%%%%%%%%
The combined distribution (Figure~\ref{fig:lat_vs_freq_dist}.c) highlights the complementarity between nKOM and nLF emissions. While peaks in both hemispheres appear connected, equatorial minima remain disconnected, especially between $30$ and $40$ kHz, forming an arc-like structure in the latitude-frequency space.
%Finally, the combined distribution of nKOM+nLF (Figure \ref{fig:lat_vs_freq_dist}.c) suggests that the nKOM and nLF distributions are complementary. The nKOM and nLF occurrence peaks distributed in the North and South appear connected. However, the nKOM and nLF occurrence minima around the equator seem disconnected between $30$~kHz and $40$~kHz. Moreover, the absence of nKOM and nLF occurrences between the equatorial occurrence minima and the Northern occurrence peaks is more pronounced and appears to follow an arc-like curvature in the latitude-frequency plane.
%%%%%%%%%%%%%%%%%%%%%%%%%%%%%%%%%%%%%%%%%%%%%%%%%%%%%%%%%%%%%%%%%%%%%%%%%%%%%

%%%%%%%%%%%%%%%%%%%%%%%%%%%%%%%%%%%%%%%%%%%%%%%%%%%%%%%%%%%%%%%%%%%%%%%%%%%%%
%_nLF => radio sources dans l'IPT ?
The nKOM and nLF radio components share similar time-frequency morphology. Furthermore, the nKOM and nLF radio components seem complementary in the latitude-frequency plane. In addition, Juno/Waves direction findings measurements on two nLF events have found their radio sources to be distributed at the outer edge of the IPT \cite{Imai2017}. Thus, the Juno/Waves observations suggest that the nLF also has its radio sources located near the IPT.

%%%%%%%%%%%%%%%%%%%%%%%%%%%%%%%%%%%%%%%%%%%%%%%%%%%%%%%%%%%%%%%%%%%%%%%%%%%%%
\subsection{Propagation Modes}\label{sec:juno_data:propagation_modes}
%%%%%%%%%%%%%%%%%%%%%%%%%%%%%%%%%%%%%%%%%%%%%%%%%%%%%%%%%%%%%%%%%%%%%%%%%%%%%
%_Propriétés dispersives du plasma [annexe]
Because of the plasma dispersive-properties, there are radio waves that cannot escape the IPT. The trapped plasma normal-modes are the Whislter mode (W-mode) and the low-frequency branch of the extraordinary mode (Z-mode). The escaping plasma modes are the ordinary mode (O-mode) and the high-frequency branch of the extraordinary mode (X-mode). Because of their resonant properties, the trapped modes are also called the ``natural modes'' of the plasma and have generation mechanisms that are different from the escaping modes. 

%%%%%%%%%%%%%%%%%%%%%%%%%%%%%%%%%%%%%%%%%%%%%%%%%%%%%%%%%%%%%%%%%%%%%%%%%%%%%
%_Juno/Waves observe nKOM & nLF dans le plasma
We want to characterize the nKOM and nLF depending on their propagation mode. In order to characterize the mode linked to an observed radio wave, the polarization of the measured signal (i.e., the direction of the inbound electric field) and the medium characteristics (i.e., the plasma density and the magnetic field properties) must be determined. However, the Juno/Waves antennas cannot provide data on wave polarization. Nevertheless, Juno's orbit allows Waves observation inside the IPT $\sim$2 days before the perijoves.

%%%%%%%%%%%%%%%%%%%%%%%%%%%%%%%%%%%%%%%%%%%%%%%%%%%%%%%%%%%%%%%%%%%%%%%%%%%%%
%_Juno/JADE fpe (JADE; McComas 2017) + Juno/FGM fce => on calcule les fréquences de coupures
%_Juno/JADE incomplet => interpolation linéaire 1min
Using the Juno JADE estimation of the cold electron density derived from measurement of the electron pitch angle \cite{McComas2017} and the Juno-MAG magnetic field measurements \cite{Connerney2017}, we are able to estimate locally to Juno the value of $\omega_{pe}$ and $\omega_{ce}$. The Juno-JADE and Juno-MAG data are resampled at 1min to match the Juno/Waves calibrated flux density data \cite{Louis2021, Louis2021_catalog}. There are gaps in the JADE measurements, resulting in either absence of pitch angle measurements, or because the plasma is too thin in comparison to the JADE instrument sensitivity. In the latter case, we choose to fill the missing data by 0. In the former case, we choose to fill the missing data by linear interpolation.

%%%%%%%%%%%%%%%%%%%%%%%%%%%%%%%%%%%%%%%%%%%%%%%%%%%%%%%%%%%%%%%%%%%%%%%%%%%%%
%_On peut séparer les obs. nKOM, nLF en plusieurs catégories: W, ZW*, Z, ZO, O et XO* 
%_On ne peut observer que des modes piégés dans les régions W, ZW* et Z:
%   => cela ne nous intéresse pas de les étudier séparément =>on va regrouper W + ZW* + Z => ZW-mode
%_On ne peut observer que des modes échappés dans les régions O et XO*:
%   => la région O pourrait se montrer intéressant => en pratique, pour les obs. Juno, cela ne concerne qu'un faible échantillon d'obs. pas exploitable avec notre méthode
%   => on va regrouper O + XO* => XO-mode
%_On peut observer du Z-mode piégé et O-mode échappé dans la région ZO:
%   => incapable de séparer les deux avec les fréquences de coupures
%_Figure [dist_ZW_ZO_XO] représente les dist en lat. vs. freq. pour les nKOM et nLF observés dans les régions ZW, ZO et XO. Les distributions pour les modes W, ZW*, Z, ZO, O et XO* sont affichés dans l'ANNEXE [dist_ZW_ZO_XO].

The JADE $\omega_{pe}$ and FGM $\omega_{ce}$ estimations are used to identify the frequency range accessible to different wave modes during Juno's passage through the plasma (see \ref{appendix:dispersion}). The Juno/Waves nKOM and nLF observations were classified in three groups depending on their observation frequency $f$ with $\omega = 2\pi f$: 

\begin{enumerate}
    \item the trapped emissions: Z- and W-modes (noted ZW-mode), if $f < f_{pe}$
    \item the escaped emissions: X- and O-modes (noted XO-mode), if $f > f_{uh}$
    \item the undeterminable emissions: Z- and O-modes (noted ZO-mode), if $f_{pe} < f < f_{uh}$
\end{enumerate}

%%%%%%%%%%%%%%%%%%%%%%%%%%%%%%%%%%%%%%%%%%%%%%%%%%%%%%%%%%%%%%%%%%%%%%%%%%%%%

\begin{figure}[!ht]
\centering
\noindent\includegraphics[width=\textwidth]{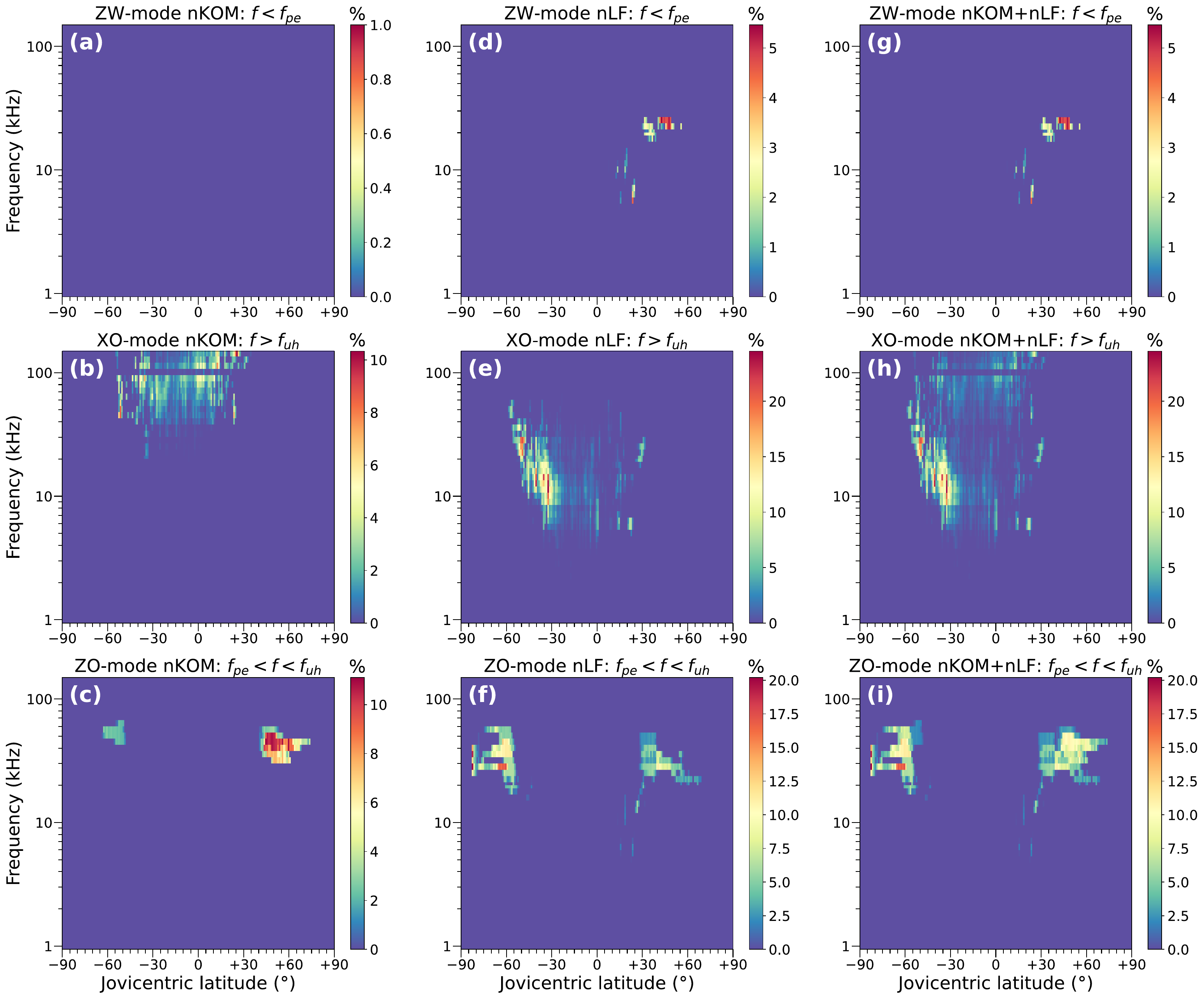}
\caption{nKOM, nLF and nKOM+nLF occurrence probability versus frequency and latitude in ZW-, XO, and ZO-modes. \change{}{The color scale is adjusted in each panel and represents the occurrence probability in \%. }Latitude bins are 1° wide and frequencies correspond to Juno/Waves frequency channels from 10 kHz to 141 kHz, distributed on a logarithmic scale with a ratio $\times 1.12$ between consecutive channels. (a) nKOM in ZW-mode, (b) nKOM in XO-mode, (c) nKOM in ZO-mode, (d) nLF in ZW-mode, (e) nLF in XO-mode, (f) nLF in ZO-mode, (g) nKOM+nLF in ZW-mode, (h) nKOM+nLF in XO-mode and (i) nKOM+nLF in ZO-mode.}
\label{fig:lat_vs_freq_dist:modes}
\end{figure}

%%%%%%%%%%%%%%%%%%%%%%%%%%%%%%%%%%%%%%%%%%%%%%%%%%%%%%%%%%%%%%%%%%%%%%%%%%%%%
%_distribution
%   (a) ZW-mode: 
%       i) pas de nKOM 
%       ii) un peu de nLF: <5%
%       => nKOM & nLF observé majoritairement non piégés
%   (b) XO-mode:
%       i) nKOM partie "HF" et "basse-latitude": >20 kHz et >-60° et <+50°
%       ii) nLF partie "BF" et "basse-latitude": <50 kHz et >-60° et <+50°
%       => logique: principalement les observations quand Juno hors IPT et loin Jupiter.
%   (c) ZO-mode:
%       i) nKOM partie "BF" et "hautes-latitude": dans 30 kHz--60 kHz et <-60° et >+50°
%       i) nLF partie "HF" et "hautes-latitude": dans 10 kHz--60 kHz et <-60° et >+50°
%       => logique: correspond aux observations proches des PJs où fce>fpe
%_A noter que: 
%   1) XO-mode nKOM+nLF discontinue
%   2) ZO-mode nKOM+nLF + XO-mode nLF continue (sauf dans nord ?)
%   => Characteristiques différentes ?
Figure \ref{fig:lat_vs_freq_dist:modes} displays the occurrence probability of nKOM, nLF and nKOM+nLF in ZW-, XO- and ZO-modes with respect to the latitude and frequency of the Juno/Waves observations. 
%%%%%%%%%%%%%%%%%%%%%%%%%%%%%%%%%%%%%%%%%%%%%%%%%%%%%%%%%%%%%%%%%%%%%%%%%%%%%
%   (i) nKOM: 
%       (a) pas de ZW-mode 
%       (b) XO-mode partie "basse-latitudes": ~>40 kHz entre >-60° et <+50°
%       (c) ZO-mode partie "hautes-latitudes": [30 kHz, 60 kHz] et <-60° et >+50°
%       => ZO-mode semble discontinue du XO-mode dans le Nord
nKOM appears exclusively in XO- and ZO-modes. XO-mode occurrences (Figure~\ref{fig:lat_vs_freq_dist:modes}.b) are found above $40$ kHz between $-55^{\circ}$ and $+40^{\circ}$, corresponding to low-latitude emissions. ZO-mode occurrences (Figure~\ref{fig:lat_vs_freq_dist:modes}.c) are seen between $30–60$ kHz at high-latitudes ($<-50^{\circ}$ and $>+40^{\circ}$), indicating a separation between low- and high-latitude sources.
%Figure \ref{fig:lat_vs_freq_dist:modes} panel (a), shows that there is no nKOM occurrence in the ZW-mode. Figure \ref{fig:lat_vs_freq_dist:modes} panel (b), shows that nKOM occurrences in the XO-mode are mainly observed $>40$~kHz within $-55^{\circ}$ and $+40^{\circ}$. This corresponds to the "low-latitude" nKOM discussed in \citeA{Boudouma2024}. Figure \ref{fig:lat_vs_freq_dist:modes} panel (c), show that nKOM occurrences in the ZO-mode are observed at 30~kHz--60~kHz outwith $-50^{\circ}$ and $+40^{\circ}$. This corresponds to the "high-latitude" nKOM discussed in \citeA{Boudouma2024}.
%%%%%%%%%%%%%%%%%%%%%%%%%%%%%%%%%%%%%%%%%%%%%%%%%%%%%%%%%%%%%%%%%%%%%%%%%%%%%
%   (ii) nLF: 
%       (d) ZW-mode [+10°, +55°]
%       (e) XO-mode: ~>40 kHz entre >-60° et <+50°
%       (f) ZO-mode: [30 kHz, 60 kHz] et <-60° et >+50°
%       => ZO-mode semble discontinue du XO-mode dans le Nord
nLF is observed in all three modes. ZW-mode occurrences (Figure~\ref{fig:lat_vs_freq_dist:modes}.d) appear below $30$ kHz in the Northern hemisphere ($+10^{\circ}$ to $+60^{\circ}$). XO-mode events (Figure~\ref{fig:lat_vs_freq_dist:modes}.e) occur over a broad latitude range ($-55^{\circ}$ to $+35^{\circ}$) above $5$ kHz. ZO-mode events (Figure~\ref{fig:lat_vs_freq_dist:modes}.f) are concentrated between $20–50$ kHz, with a notable low-frequency extension down to $7$ kHz in the North.
%Figure \ref{fig:lat_vs_freq_dist:modes} panel (d), shows that there is nLF occurrences in ZW-mode $<$30~kHz in the North within $+10^{\circ}$ and $+60^{\circ}$. Figure \ref{fig:lat_vs_freq_dist:modes} panel (e), shows that nLF occurrences in the XO-mode are mainly observed $>5$~kHz within $-55^{\circ}$ and $+35^{\circ}$. Figure \ref{fig:lat_vs_freq_dist:modes} panel (f), shows that nLF occurrences in the ZO-mode are mainly observed at 20~kHz--50~kHz within $-55^{\circ}$ and $+30^{\circ}$. The Northern peak of nLF occurrence in ZO-mode presents a low-frequency extension down to 7~kHz within $+15^{\circ}$ and $+30^{\circ}$.
%%%%%%%%%%%%%%%%%%%%%%%%%%%%%%%%%%%%%%%%%%%%%%%%%%%%%%%%%%%%%%%%%%%%%%%%%%%%%
%   (iii) nKOM+nLF:
%       (h) ZW-mode [+10°, +55°] == nLF (bien-sûr)
%       (e) XO-mode: ~>40 kHz entre >-60° et <+50°
%       (f) ZO-mode: [30 kHz, 60 kHz] et <-60° et >+50°
%       => ZO-mode semble discontinue du XO-mode dans le Nord
The combined distribution confirms these patterns. ZW-mode (Figure \ref{fig:lat_vs_freq_dist:modes}.g) is identical to the nLF distribution (Figure \ref{fig:lat_vs_freq_dist:modes}.d). In XO-mode (Figure \ref{fig:lat_vs_freq_dist:modes}.h), a gap appears at $30–40$ kHz near the equator, separating nKOM and nLF. In ZO-mode (Figure \ref{fig:lat_vs_freq_dist:modes}.i), the Northern and Southern peaks of both components show continuity across latitude.
%Figure \ref{fig:lat_vs_freq_dist:modes} panel (g) is identical to Figure \ref{fig:lat_vs_freq_dist:modes} panel (d), as (a) is empty. Figure \ref{fig:lat_vs_freq_dist:modes} panel (h), shows that there is a discontinuity between nKOM and nLF occurrences in XO-mode at 30~kHz--40~kHz within $-30^{\circ}$ and $+30^{\circ}$. Figure \ref{fig:lat_vs_freq_dist:modes} panel (i), shows that there is a continuity between Northern and Southern peaks of nKOM and nLF.

%%%%%%%%%%%%%%%%%%%%%%%%%%%%%%%%%%%%%%%%%%%%%%%%%%%%%%%%%%%%%%%%%%%%%%%%%%%%%
%_On étudie la génération de nKOM et nLF en X- et O-modes:
%   => on exclus les observations ZW-mode : ça ne concerne que le nLF
As this study focuses on the generation of the nKOM and nLF in the X- and O-modes, we propose to exclude the observations in ZW-mode from the dataset. Nevertheless, only nLF has been observed in ZW-mode. In comparison to the observations in XO- and ZO-modes, the observations in ZW-mode represent a very small subset of the total observations: their removal from the dataset does not change the overall structure of the distribution.

\section{LsPRESSO: Modeling of the Occurrence Probability Distribution versus the Latitude and Frequency}\label{sec:LsPRESSO}
As discussed in B24, the nKOM and nLF occurrence probability distributions versus latitude and frequency are assumed to be representative of the emissions time-averaged latitudinal beaming at large scale. We also assume that all the potential nKOM and nLF radio sources are permanently active and emitting under the specific conditions described Section \ref{sec:LsPRESSO:params}.
%%%%%%%%%%%%%%%%%%%%%%%%%%%%%%%%%%%%%%%%%%%%%%%%%%%%%%%%%%%%%%%%%%%%%%%%%%%%%

%%%%%%%%%%%%%%%%%%%%%%%%%%%%%%%%%%%%%%%%%%%%%%%%%%%%%%%%%%%%%%%%%%%%%%%%%%%%%
We use the \citeA{Imai2016} diffusive density model and \citeA{Connerney1998} VIP4 magnetic field model to compute the cold electron density $n_e$ and magnetic field vector $\mathbf{B}$ at any point of Jupiter's inner magnetosphere. The 3D-Cartesian grid used is identical as in B24, i.e., $-15 \leq x,y \leq +15$ $R_J$ in the equatorial plane and $-10 \leq z \leq +10$ $R_J$ along the rotation axis. The grid step is $dl~=~0.1$ $R_J$.
\subsection{Large-Scale Generation Scenarios and Parameters}\label{sec:LsPRESSO:params}
%%%%%%%%%%%%%%%%%%%%%%%%%%%%%%%%%%%%%%%%%%%%%%%%%%%%%%%%%%%%%%%%%%%%%%%%%%%%%
%_LsPRESSO: simulation de l'observation d'émissions plasma à grande échelle
LsPRESSO is used to simulate the Juno/Waves observations of escaping radio waves produced inside the IPT at large-scale. 
%%%%%%%%%%%%%%%%%%%%%%%%%%%%%%%%%%%%%%%%%%%%%%%%%%%%%%%%%%%%%%%%%%%%%%%%%%%%%
%_Scénarios de générations:
%   1) f=fpe, r||-grad(ne) => lineaire et nonlineaire
%   2) f=2pe, r||-grad(ne) => nonlineaire
As proposed in B24, two generation scenarios can be constructed based on the theoretical models discussed in Section \ref{sec:mode_conversion}:
\begin{enumerate}
    \item[\textbullet] Scenario \#1, based on the theory of \citeA{Jones1980}: Generation of Jovian plasma emissions at the fundamental plasma frequency, $\omega \sim \omega_{pe}$, in the directions $\mathbf{r}+$ and $\mathbf{r}-$, such that $\angle(\mathbf{r}{\pm}, \mathbf{B}) = \frac{\pi}{2} \mp (\frac{\pi}{2} - \beta)$, with $\beta = \arctan (\sqrt{\omega_{pe}/\omega_{ce}})$.
    \item[\textbullet] Scenario \#2, based on the theory of \citeA{Fung&Papadopoulos1987}: Generation of Jovian plasma emissions at the first harmonic of the upper-hybrid frequency, $\omega \sim 2\omega_{uh}$, in all directions $\mathbf{r} \perp \mathbf{B}$.
\end{enumerate}

%%%%%%%%%%%%%%%%%%%%%%%%%%%%%%%%%%%%%%%%%%%%%%%%%%%%%%%%%%%%%%%%%%%%%%%%%%%%%
%_Snell-Descartes : r || -grad(ne)
%_A grande échelle:
%   _directivité des émissions à fpe => car produit proche du cutoff (grad(N) très fort)
%   _directivité des émissions à 2fpe quand grad(ne) fort-

The Snell-Descartes law shows that the waves beaming tends to align with the normal to the refractive index iso-surfaces if this index varies quickly. The resulting ray is aligned with $-\nabla n_e$ in O-mode but also in X-mode (if $\omega_{pe} \gg \omega_{ce}$). At large-scale, such beaming occurs for radio waves produced near $\omega_{pe}$ as the refractive index varies quickly after the cutoff (cf. Figure \ref{appendix:fig:dispersion_curve}). This also can be the case for radio waves produced near $2\omega_{pe}$ if $\nabla n_e$ is strong. 
Thus, we propose two additional generation scenarios where the ray direction is $\mathbf{r} \parallel -\nabla n_e$:
\begin{itemize}
    \item[\textbullet] Scenario \#3: near the fundamental $f \sim \omega_{pe}$, consistent with linear and nonlinear conversion mode mechanisms.
    \item[\textbullet] Scenario \#4: near the first harmonic $f \sim 2\omega_{pe}$, consistent with nonlinear conversion mode mechanisms.
\end{itemize}
%%%%%%%%%%%%%%%%%%%%%%%%%%%%%%%%%%%%%%%%%%%%%%%%%%%%%%%%%%%%%%%%%%%%%%%%%%%%%

%%%%%%%%%%%%%%%%%%%%%%%%%%%%%%%%%%%%%%%%%%%%%%%%%%%%%%%%%%%%%%%%%%%%%%%%%%%%%
%_D'après section 2:
%   1) Lineaire: Te, alpha, L et B0:
%   2) Nonlinear: Te, nb/n0, L et B0:
As seen in the Section \ref{sec:mode_conversion}, LMC mechanisms are mainly controlled by the electron temperature and the angle between the density gradient and the magnetic field. In contrast, NLMC mechanisms are primarily constrained by the Langmuir waves amplitude, the temperature ratio between ion and electron species, and small-scale plasma density fluctuations.
%%%%%%%%%%%%%%%%%%%%%%%%%%%%%%%%%%%%%%%%%%%%%%%%%%%%%%%%%%%%%%%%%%%%%%%%%%%%%
%   => Or à grande échelle dx ~10^5*lambda: 
%       A) on n'a pas accès à Te et nb/n0
%       B) grad(ne) et B pas représentatif de L et B0 car grande échelle
However, the \citeA{Imai2016} diffusive density model does not allow for the calculation of the plasma electron temperature, making it difficult to use for estimating the Langmuir waves amplitude or the density fluctuations order of magnitude.
%%%%%%%%%%%%%%%%%%%%%%%%%%%%%%%%%%%%%%%%%%%%%%%%%%%%%%%%%%%%%%%%%%%%%%%%%%%%%

%%%%%%%%%%%%%%%%%%%%%%%%%%%%%%%%%%%%%%%%%%%%%%%%%%%%%%%%%%%%%%%%%%%%%%%%%%%%%
%_On propose 2 paramètres:
%   1) alpha: configuration géometrique marcrospique entre grad(ne) et B
%       _Raison 1: théorique: 
%           A) efficacité conversion linéaire
%           B) pourquoi pas impliqué dans l'efficacité de la conversion nonlineaire (cf. Fung)
%       _Raison 2: observationnelle:
%           A) génération de nKOM et nLF contrôlée par la reconfiguration de la magnetosphere
%   2) epsilon: force de grad(ne) à grande échelle
%       _Raison 1: théorique:
%           A) efficacité lineaire et nonlinear contrôlé par L => grad(ne) != L mais doivent être liés
%       _Raison 2: géometrique
%           A) r || -grad(ne) valable quand grad(ne) fort => vérifier la validité de l'hypothèse cas 2fpe
%%%%%%%%%%%%%%%%%%%%%%%%%%%%%%%%%%%%%%%%%%%%%%%%%%%%%%%%%%%%%%%%%%%%%%%%%%%%%
Thus, we choose to constrain the activation of sources with the same parameters as in B24:
\begin{enumerate}
    \item[\textbullet] The angle between the large-scale density gradient and the magnetic field vector, $\angle({\boldsymbol\nabla} n_e, \mathbf{B})$. This parameter is related to LMC mechanisms and potentially to NLMC mechanisms (as in the theory of \citeA{Fung&Papadopoulos1987}).
    \item[\textbullet] The large-scale density gradient magnitude, $||{\boldsymbol\nabla} n_e||$. This parameter is associated with both LMC and NLMC mechanisms. Additionally, it accounts for the large-scale plasma dispersive effects and helps to constrain the validity of the assumptions proposed by generation scenarios regarding the beam directivity.
\end{enumerate}

%%%%%%%%%%%%%%%%%%%%%%%%%%%%%%%%%%%%%%%%%%%%%%%%%%%%%%%%%%%%%%%%%%%%%%%%%%%%%
\subsection{Results}\label{sec:results}
%%%%%%%%%%%%%%%%%%%%%%%%%%%%%%%%%%%%%%%%%%%%%%%%%%%%%%%%%%%%%%%%%%%%%%%%%%%%%

Table \ref{tab:scenario} summarizes the generation scenarios proposed for simulating Jovian plasma emissions. 

\begin{table}[!h]
\centering
    \caption{Generation scenarios of Jovian plasma emissions simulated with LsPRESSO.}
    \begin{tabular}{l|c|c}
    Generation Scenario & Frequency & Directivity \\\hline
    \#1: LMC Jones (1980) & $\omega_{pe}$ & $\angle(\mathbf{r}{\pm}, \mathbf{B}) = \frac{\pi}{2} \mp (\frac{\pi}{2} - \beta)$ \\
    \#2: NLMC Fung \& Papadopoulos (1987) & $2\omega{uh}$ & $\angle(\mathbf{r}, \mathbf{B}) = \frac{\pi}{2}$ \\
    \#3: Fundamental Plasma Frequency & $\omega_{pe}$ & $\angle(\mathbf{r}, {\boldsymbol\nabla} n_e) = \pi$ \\
    \#4: Harmonic Plasma Frequency & $2\omega_{pe}$ & $\angle(\mathbf{r}, {\boldsymbol\nabla} n_e) = \pi$ \\
    \end{tabular}
    
    \label{tab:scenario}
\end{table}

The activation condition of radio sources is constrained by the parameter pair $\alpha$ and $\epsilon$. The parameter $\alpha$ represents the angle $\angle({\boldsymbol\nabla} n_e, \mathbf{B})$ and varies within the range $\alpha\in[0^{\circ},90^{\circ}]$ in steps of $3^{\circ}$. The parameter $\epsilon$ is defined as the \change{centile}{percentile} of $||{\boldsymbol\nabla} n_e||$ at each observation frequency, varying within $\epsilon\in[0\%, 100\%]$ in steps of $10\%$. For each generation scenario and each pair $(\alpha,\epsilon)$, a time-frequency visibility map is modeled by computing the visibility of emissions at all positions and observation frequencies of Juno/Waves. Their compatibility with the nKOM and nLF observations is deduced by comparing the modeled latitude and frequency distribution to the observed ones using the method introduced in B24\change{}{, i.e, by retrieving the set of parameters $(\alpha,\epsilon)$ that maximizes the linear correlation coefficient $C$ between the observed and modeled distributions. The complete step-by-step process is fully detailed in B24 at the end of Section 4.2}.
%%%%%%%%%%%%%%%%%%%%%%%%%%%%%%%%%%%%%%%%%%%%%%%%%%%%%%%%%%%%%%%%%%%%%%%%%%%%%

%%%%%%%%%%%%%%%%%%%%%%%%%%%%%%%%%%%%%%%%%%%%%%%%%%%%%%%%%%%%%%%%%%%%%%%%%%%%%
Table \ref{tab:results} summarizes, for each generation scenario, the maximum value of the linear correlation coefficient $C_{max}$ obtained between the modeled distribution and the observed distributions of nKOM, nLF, and nKOM+nLF observed by Juno/Waves, independently of the mode, in XO and in ZO. The Figures summarizing the results obtained for all distributions observed by Juno/Waves are presented in the Supplementary Informations.

%Results - Correlation
\begin{table}[!h]
    \centering
    \caption{Maximum value of the linear correlation coefficient $C_{max}$ between the simulated distributions for each generation scenario and the observed distributions of nKOM, nLF, and nKOM+nLF measured by Juno/Waves, without mode distinction (i.e., global distribution), in XO and in ZO. Values of $C_{max} > 35\%$ are highlighted in bold.}
    \begin{tabular}{| c | c | c | c | c | c | c | c | c | c | c |}
        \hline
        \multirow{3}{*}{Cutoff} & \multirow{3}{*}{Scenario} & \multicolumn{9}{c|}{Correlation Maximum $C_{max}$} \\\cline{3-11}
        & & \multicolumn{3}{c|}{nKOM} & \multicolumn{3}{c|}{nLF}  & \multicolumn{3}{c|}{nKOM+nLF}  \\\cline{3-11}
        & & Global & XO & ZO & Global & XO & ZO & Global & XO & ZO \\
        \hline
        %MODE O
        \multirow{4}{*}{O-mode} & \#1 & 
        21\% & 26\% & 22\% & %nKOM
        -2\% & 5\% & 9\% & %nLF
        8\% & 1\% & 12\%   %nKOM+nLF
        \\\cline{2-11}
        & \#2 & 
        23\% & \textbf{51\%} & 0\% & %nKOM
        1\% & 2\% & -1\% & %nLF
        13\% & 24\% & -1\%   %nKOM+nLF
        \\\cline{2-11}
        & \#3 & 
        \textbf{39\%} & 27\% & \textbf{37\%} & %nKOM %37% is 10-141 kHz
        \textbf{41\%} & \textbf{48\%} & 22\% & %nLF
        \textbf{44\%} & \textbf{46\%} & 29\%   %nKOM+nLF
        \\\cline{2-11}
        & \#4 & 
        22\% & 25\% & 20\% & %nKOM
        \textbf{41\%} & \textbf{45\%} & \textbf{55\%} & %nLF
        \textbf{45\%} & \textbf{41\%} & \textbf{54\%}   %nKOM+nLF
        \\\hline
        %MODE X
        \multirow{4}{*}{X-mode} & \#1 & 
        11\% & 20\% & 0\% & %nKOM
        -1\% & 0\% & 9\% & %nLF
        4\% & 6\% & 9\%   %nKOM+nLF
        \\\cline{2-11}
        & \#2 & 
        22\% & \textbf{50\%} & 0\% & %nKOM
        1\% & 2\% & -1\% & %nLF
        12\% & 21\% & -1\%   %nKOM+nLF
        \\\cline{2-11}
        & \#3 & 
        24\% & \textbf{47\%} & 0\% & %nKOM
        0\% & 0\% & 0\% & %nLF
        7\% & 13\% & 0\%   %nKOM+nLF
        \\\cline{2-11}
        & \#4& 
        26\% & \textbf{53\%} & 0\% & %nKOM
        \textbf{38\%} & \textbf{49\%} & 0\% & %nLF
        33\% & \textbf{46\%} & 0\%   %nKOM+nLF
        \\\hline
        %\multirow{4}{*}{mode O} 
        %& Scénario \#1 & \multicolumn{3}{*}{20\% & 25\% & 20\% }  & &\\
        %& Scénario \#2 & 22\% & &\\
        %& Scénario \#3 & 39\% & &\\
        %& Scénario \#4 & 15\% & &\\ 
    \end{tabular}
    \label{tab:results}
\end{table}

%%%%%%%%%%%%%%%%%%%%%%%%%%%%%%%%%%%%%%%%%%%%%%%%%%%%%%%%%%%%%%%%%%%%%%%%%%%%%
%_3) Best results
%   nKOM:
%       total : [O mode: #3]
For nKOM, the results show that only scenario \#3 in O-mode is compatible with the global distribution, with $C_{max}~=~39\%$.
%       XO-mode : [O mode: #2] and [X mode: #2, #3, #4]
Scenario \#2 in O-mode and scenarios \#2, \#3, and \#4 in X-mode are compatible with the XO distribution, with $C_{max}~=~51\%$, $50\%$, $47\%$, and $53\%$, respectively.
%       ZO-mode : [O mode: #3]
Only scenario \#3 in O-mode is compatible with the ZO distribution, with $C_{max}~=~37\%$.
%%%%%%%%%%%%%%%%%%%%%%%%%%%%%%%%%%%%%%%%%%%%%%%%%%%%%%%%%%%%%%%%%%%%%%%%%%%%%

%%%%%%%%%%%%%%%%%%%%%%%%%%%%%%%%%%%%%%%%%%%%%%%%%%%%%%%%%%%%%%%%%%%%%%%%%%%%%
%   nLF:
%       total : [O mode: #3, #4] and [X mode: #4]
For nLF, the results show that scenarios \#3 and \#4 in O-mode and scenario \#4 in X-mode are compatible with the global distribution, with $C_{max}~=~41\%$, $41\%$, and $38\%$, respectively.
%       XO-mode : [O mode: #3, #4] and [X mode: #4]
Similarly, scenarios \#3 and \#4 in O-mode and scenario \#4 in X-mode are compatible with the XO distribution, with $C_{max}~=~48\%$, $45\%$, and $49\%$, respectively.
%       ZO-mode : [O mode: #4]
However, only scenario \#4 in O-mode is compatible with the ZO distribution, with $C_{max}~=~55\%$.
%%%%%%%%%%%%%%%%%%%%%%%%%%%%%%%%%%%%%%%%%%%%%%%%%%%%%%%%%%%%%%%%%%%%%%%%%%%%%

%%%%%%%%%%%%%%%%%%%%%%%%%%%%%%%%%%%%%%%%%%%%%%%%%%%%%%%%%%%%%%%%%%%%%%%%%%%%%
%   nKOM+nLF:
%       total : [O mode: #3, #4]
For nKOM+nLF, the results show that scenarios \#3 and \#4 in O-mode are compatible with the global distribution, with $C_{max}~=~44\%$ and $45\%$, respectively.
%       XO-mode : [O mode: #3, #4] and [X mode: #4]
As for nLF, scenarios \#3 and \#4 in O-mode and scenario \#4 in X-mode are compatible with the XO distribution, with $C_{max}~=~46\%$, $41\%$, and $46\%$, respectively. However, the simulated distributions correlate better with nLF alone ($C_{max}~=~48\%$, $45\%$, and $49\%$).
%       ZO-mode : [O mode: #3]
As for nLF, only scenario \#4 in O-mode is compatible with the ZO distribution, with $C_{max}~=~54\%$. The simulated distribution is nearly as well correlated with nLF alone ($C_{max}~=~55\%$).
%%%%%%%%%%%%%%%%%%%%%%%%%%%%%%%%%%%%%%%%%%%%%%%%%%%%%%%%%%%%%%%%%%%%%%%%%%%%%

%%%%%%%%%%%%%%%%%%%%%%%%%%%%%%%%%%%%%%%%%%%%%%%%%%%%%%%%%%%%%%%%%%%%%%%%%%%%%
%_4) General conclusions:
%   _Scenario #1 is incompatible with Juno/Waves observations
%   _Scenario #2 is potentially compatible with nKOM
%   _Scenarios #3 and #4 perform best (compatible with nKOM and nLF)
Thus, we can summarize that:
\begin{enumerate}
    \item[\textbullet] Scenario \#1 is incompatible with nKOM and nLF observations. This incompatibility appears to be caused by the directivity predicted by \citeA{Jones1980}.
    \item[\textbullet] Scenario \#2 is only compatible with nKOM observations in XO mode. This is expected since (i) ZO observations imply $\omega<\omega_{uh}$ and (ii) nLF observations in XO-mode are mainly distributed at frequencies too low to correspond to $2\omega_{uh}$.
    \item[\textbullet] Scenarios \#3 and \#4 are both compatible with nKOM and nLF observations. nKOM in ZO is consistent with emissions generated at $\omega_{pe}$ for oblique $\alpha$ ($\sim56^{\circ}$) and high $\epsilon$ ($\sim80\%$). nLF in XO is compatible with emissions generated in O-mode at $\omega_{pe}$ and $2\omega_{pe}$ and in X-mode at $2\omega_{pe}$ for near-perpendicular $\alpha$ ($>75^{\circ}$) and intermediate or low $\epsilon$ ($<70^{\circ}$). nLF in ZO is consistent with emissions generated in O-mode at $2\omega_{pe}$ for near-parallel $\alpha$ ($<15^{\circ}$) and very high $\epsilon$ ($>90\%$).
    \item[\textbullet] It is difficult to constrain the characteristics of nKOM in XO since it is compatible with scenarios \#2, \#3, and \#4. However, they seem to converge toward (i) emissions in X-mode and (ii) values of $\alpha > 30^{\circ}$. Finally, even though none of these scenarios can be excluded, scenario \#4 stands out: it has the highest correlation ($C_{max}~=~53\%$ vs. $C_{max}\sim50\%$) and is highly localized in the parameter space with $\alpha>45^{\circ}$ and $\epsilon\sim80\%$.
\end{enumerate}

\section{Comparison with the Juno/Waves Observations at PJ01}\label{sec:comparison}

%%%%%%%%%%%%%%%%%%%%%%%%%%%%%%%%%%%%%%%%%%%%%%%%%%%%%%%%%%%%%%%%%%%%%%%%%%%%%
%_) Juno/Waves PJ01: Radiogoniometry Measurements
%   => Comparison between predictions and observations
In the absence of polarization measurements, the Waves instrument does not allow for goniopolarimetry measurements of nKOM and nLF observations. However, due to Juno's high-latitude observation geometry, it was possible to perform goniometry measurements of nKOM and nLF on the day following the first perijove of Jupiter (PJ01) on 28/08/2016, using the spacecraft's rotation \cite{Imai2017}.
%%%%%%%%%%%%%%%%%%%%%%%%%%%%%%%%%%%%%%%%%%%%%%%%%%%%%%%%%%%%%%%%%%%%%%%%%%%%%

\subsection{Radio Sources Location}

\begin{figure}[!htbp]
  \centering
  \includegraphics[width=0.70\textwidth]{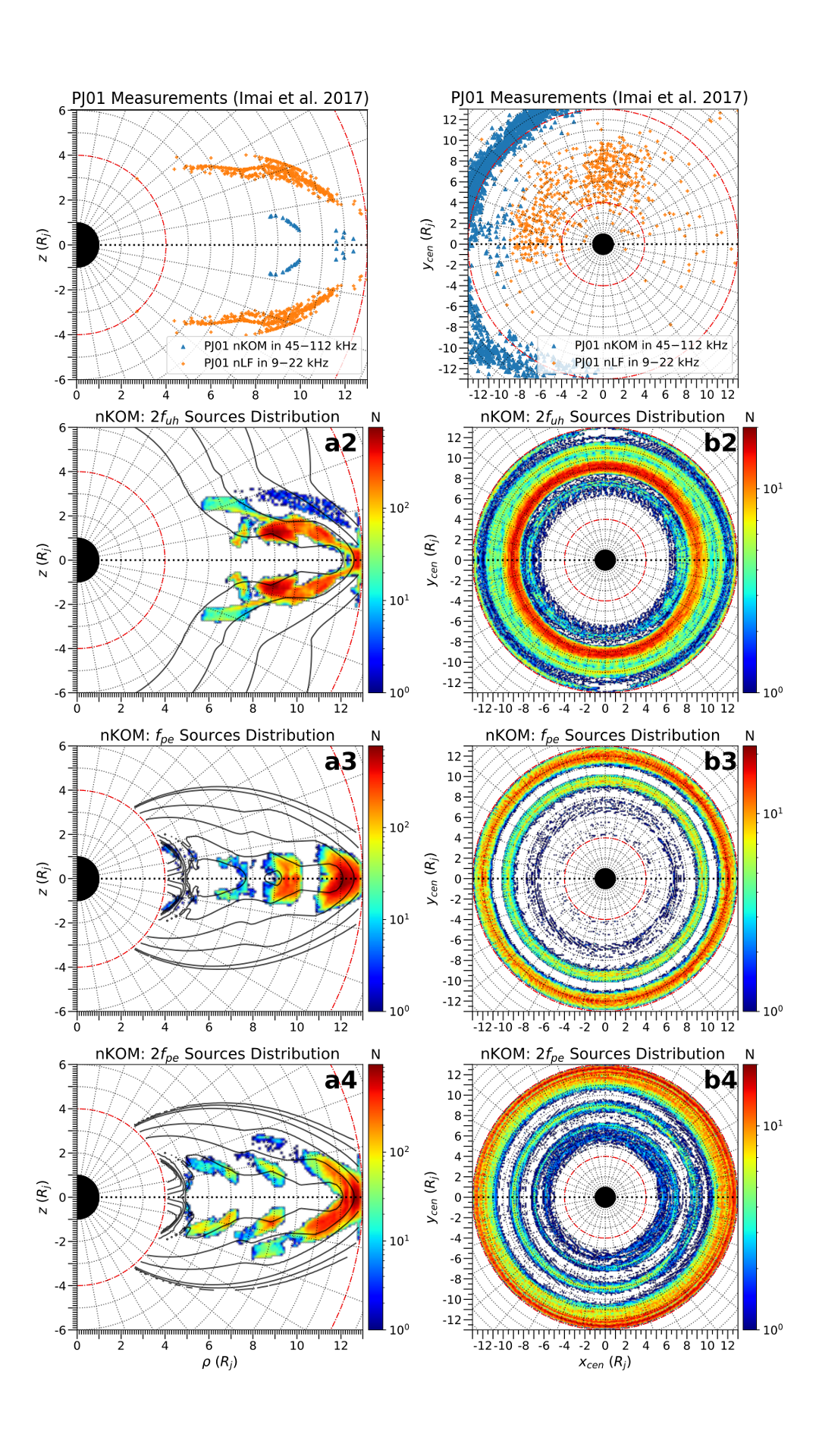}
  \caption{Meridian distributions (a2, a3, a4) and equatorial distributions (b2, b3, b4) of nKOM radio sources in the plasma disk and the Io Plasma Torus (IPT). Panels indexed 2, 3, and 4 correspond to sources predicted with LsPRESSO for the generation scenarios of the same number. Blue triangles and orange crosses indicate the positions of nKOM and nLF radio sources, respectively, as measured by Juno/Waves on 28/08/2016 and reconstructed from Figures 1a and 3 of \citeA{Imai2017}. \change{The x-axes, y-axes, and grid definitions are identical to those in Figure 3 of B24.}{The x-axes and y-axes represent respectively, for the left and right panels, the cylindrical coordinate $\rho$ and $z$ of the meridian plane and the cartesian coordinates $x_{cen}$ and $y_{cen}$ of the equatorial plane, both in the centrifugal referential. Jupiter's disk is displayed in black, the dotted circles around it mark radial distance by 1 $R_J$ steps. The dotted lines mark respectively, for the left and right panels, latitudes and CML by $10^\circ$ steps, with the bold dotted line indicating the centrifugal equator and the CML at $200.769^{\circ}$. The red dotted-dashed circles mark Imai's (2016) diffusive density model limits at 4~$R_j$ and 13~$R_j$. The meridian distribution on the top panel is mirrored around the centrifugal equator as they estimated the location of the radio sources originating from the southern hemisphere. The color scale is adjusted in each panel and represents the number of sources N. The color maps have 130$\times$120 pixels.} In panels (a2, a3, a4), the black contours correspond to the radio source frequencies $f~=~[5, 10, 20, 40, 80, 160]$~kHz (except for sources at $2\omega_{uh}$, where the first contour is at $f = 20$ kHz).}
\label{fig:sources_distribution_nKOM}
\end{figure}

\begin{figure}[!htbp]
  \centering
  \includegraphics[width=0.70\textwidth]{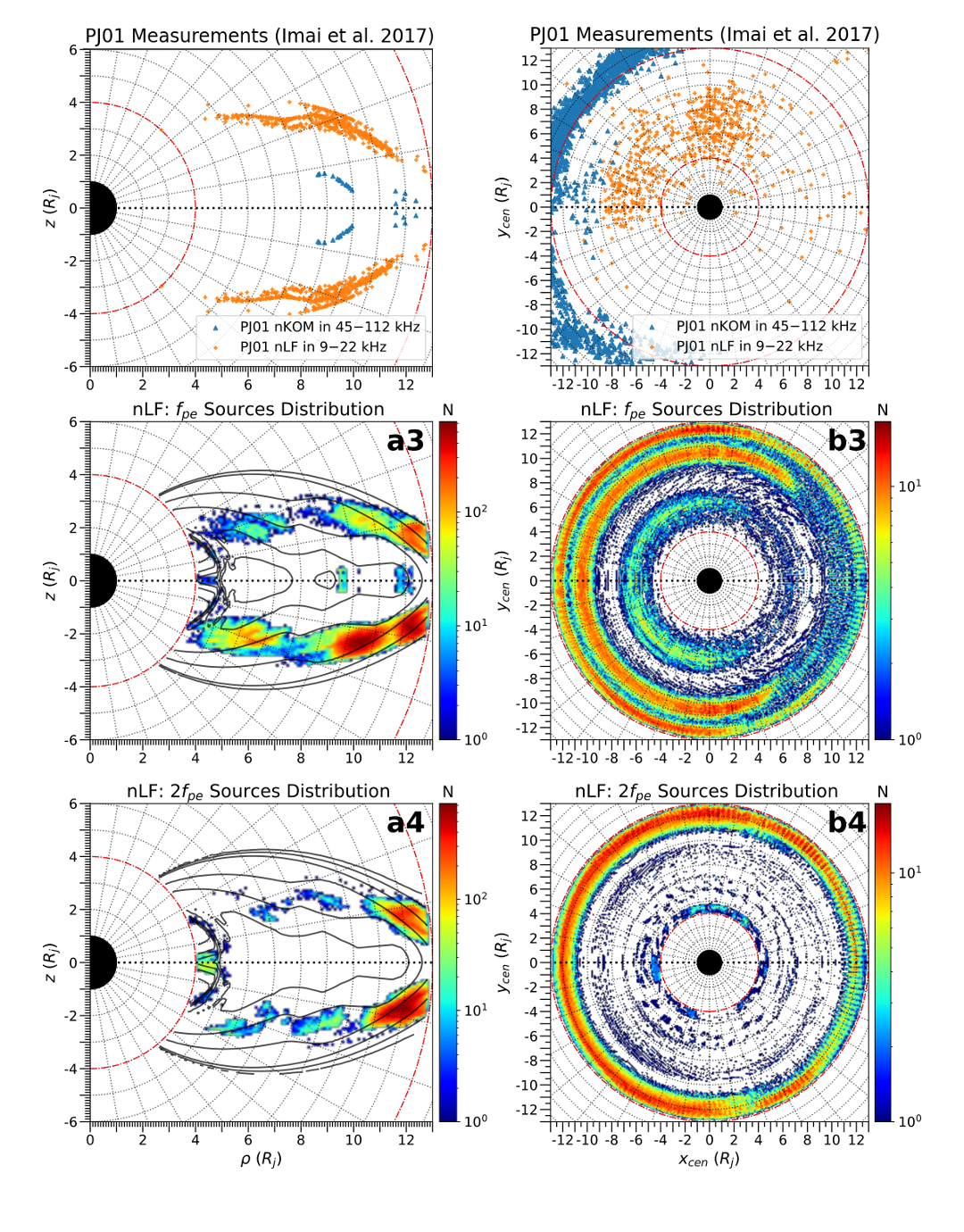}
  \caption{Meridian distributions (a3, a4) and equatorial distributions (b3, b4) of nLF radio sources in the plasma disk and the Io Plasma Torus (IPT), as measured by Juno/Waves and predicted with LsPRESSO for the compatible generation scenarios determined from Table \ref{tab:results}. Panels indexed 3 and 4 correspond to sources predicted with LsPRESSO for the generation scenarios of the same number. The panel definitions are identical to those in Figure \ref{fig:sources_distribution_nKOM}.}
\label{fig:sources_distribution_nLF}
\end{figure}

%%%%%%%%%%%%%%%%%%%%%%%%%%%%%%%%%%%%%%%%%%%%%%%%%%%%%%%%%%%%%%%%%%%%%%%%%%%%%
Figures \ref{fig:sources_distribution_nKOM} and \ref{fig:sources_distribution_nLF} show the distributions of nKOM and nLF radio sources in the plasma disk and the Io Plasma Torus (IPT), as measured by Juno/Waves during PJ01 and as predicted by LsPRESSO. In the following, we compare in details the predicted radio source locations for nKOM and nLF with those inferred from Juno/Waves observations during PJ01.
%%%%%%%%%%%%%%%%%%%%%%%%%%%%%%%%%%%%%%%%%%%%%%%%%%%%%%%%%%%%%%%%%%%%%%%%%%%%%

%%%%%%%%%%%%%%%%%%%%%%%%%%%%%%%%%%%%%%%%%%%%%%%%%%%%%%%%%%%%%%%%%%%%%%%%%%%%%

%%%%%%%%%%%%%%%%%%%%%%%%%%%%%%%%%%%%%%%%%%%%%%%%%%%%%%%%%%%%%%%%%%%%%%%%%%%%%
%_) nKOM
Panels (a) and (b) of Figure \ref{fig:sources_distribution_nKOM} show the meridian and equatorial projections, respectively, in the jovicentric reference frame of the nKOM radio sources predicted by LsPRESSO.
%%%%%%%%%%%%%%%%%%%%%%%%%%%%%%%%%%%%%%%%%%%%%%%%%%%%%%%%%%%%%%%%%%%%%%%%%%%%%
The meridian distributions of nKOM radio sources predicted with LsPRESSO partially match the Juno/Waves observations. Scenarios \#2, \#3, and \#4 (Figure \ref{fig:sources_distribution_nKOM}.a2, a3, and a4) predict nKOM radio sources between $\sim8~R_J$ and $\sim9.5~R_J$. However, only scenarios \#3 and \#4 predict nKOM radio sources near the centrifugal equator at $>11.5~R_J$. On the other hand, the equatorial distributions of nKOM radio sources predicted with LsPRESSO do not match the Juno/Waves observations. Indeed, scenarios \#2, \#3, and \#4 (Figure \ref{fig:sources_distribution_nKOM}.b2, b3, and b4) predict an azimuthal symmetry around Jupiter’s centrifugal axis in the distribution of nKOM radio sources, whereas the observed sources are located at $x_{cen}<0$.
%%%%%%%%%%%%%%%%%%%%%%%%%%%%%%%%%%%%%%%%%%%%%%%%%%%%%%%%%%%%%%%%%%%%%%%%%%%%%

%%%%%%%%%%%%%%%%%%%%%%%%%%%%%%%%%%%%%%%%%%%%%%%%%%%%%%%%%%%%%%%%%%%%%%%%%%%%%
%_) nLF
Panels (a) and (b) of Figure \ref{fig:sources_distribution_nLF} show the meridian and equatorial projections, respectively, in the jovicentric reference frame of the nLF radio sources predicted by LsPRESSO.
%%%%%%%%%%%%%%%%%%%%%%%%%%%%%%%%%%%%%%%%%%%%%%%%%%%%%%%%%%%%%%%%%%%%%%%%%%%%%
The meridian distributions of nLF radio sources predicted with LsPRESSO match the Juno/Waves observations. Scenarios \#3 and \#4 (Figure \ref{fig:sources_distribution_nLF}.a3 and a4) predict nLF radio sources at $>8.5~R_J$ on the edges of the plasma disk at latitudes of $\sim\pm5^{\circ}$. 
Although the equatorial distributions of nLF radio sources predicted with LsPRESSO only partially match the Juno/Waves observations, scenario \#3 (Figure \ref{fig:sources_distribution_nKOM}.b3) predicts the azimuthal asymmetry around Jupiter’s centrifugal axis. Finally, it is also noticeable that scenario \#3 appears to predict both the meridian and equatorial distributions of nKOM radio sources between $\sim11.5~R_J$ and $\sim12~R_J$.

%However, the equatorial distributions of nLF radio sources predicted with LsPRESSO only partially match the Juno/Waves observations. Indeed, only scenario \#3 (Figure \ref{fig:sources_distribution_nKOM}.b3) predicts the azimuthal asymmetry around Jupiter’s centrifugal axis in the distribution of nLF radio sources. Finally, it is also noticeable that scenario \#3 appears to predict both the meridian and equatorial distributions of nKOM radio sources between $\sim11.5~R_J$ and $\sim12~R_J$.
%%%%%%%%%%%%%%%%%%%%%%%%%%%%%%%%%%%%%%%%%%%%%%%%%%%%%%%%%%%%%%%%%%%%%%%%%%%%%

%%%%%%%%%%%%%%%%%%%%%%%%%%%%%%%%%%%%%%%%%%%%%%%%%%%%%%%%%%%%%%%%%%%%%%%%%%%%%
Thus, the localization of nKOM and nLF radio sources observed by Juno/Waves is partially consistent with the predictions of LsPRESSO. The LsPRESSO predictions for nLF appear to explain the localization of both nKOM and nLF radio sources. These sources correspond to Jovian plasma emissions generated at the fundamental frequency $\omega_{pe}$ for an O-mode cutoff, with radio sources where ${\boldsymbol\nabla} n_e \perp \mathbf{B}$. This generation scenario is the only one capable of reproducing the asymmetry in the longitudinal distribution of nKOM and nLF radio sources.
%%%%%%%%%%%%%%%%%%%%%%%%%%%%%%%%%%%%%%%%%%%%%%%%%%%%%%%%%%%%%%%%%%%%%%%%%%%%%

\subsection{Time-Frequency Morphology}

\begin{figure}[htbp]
  \centering
  \includegraphics[width=\textwidth]{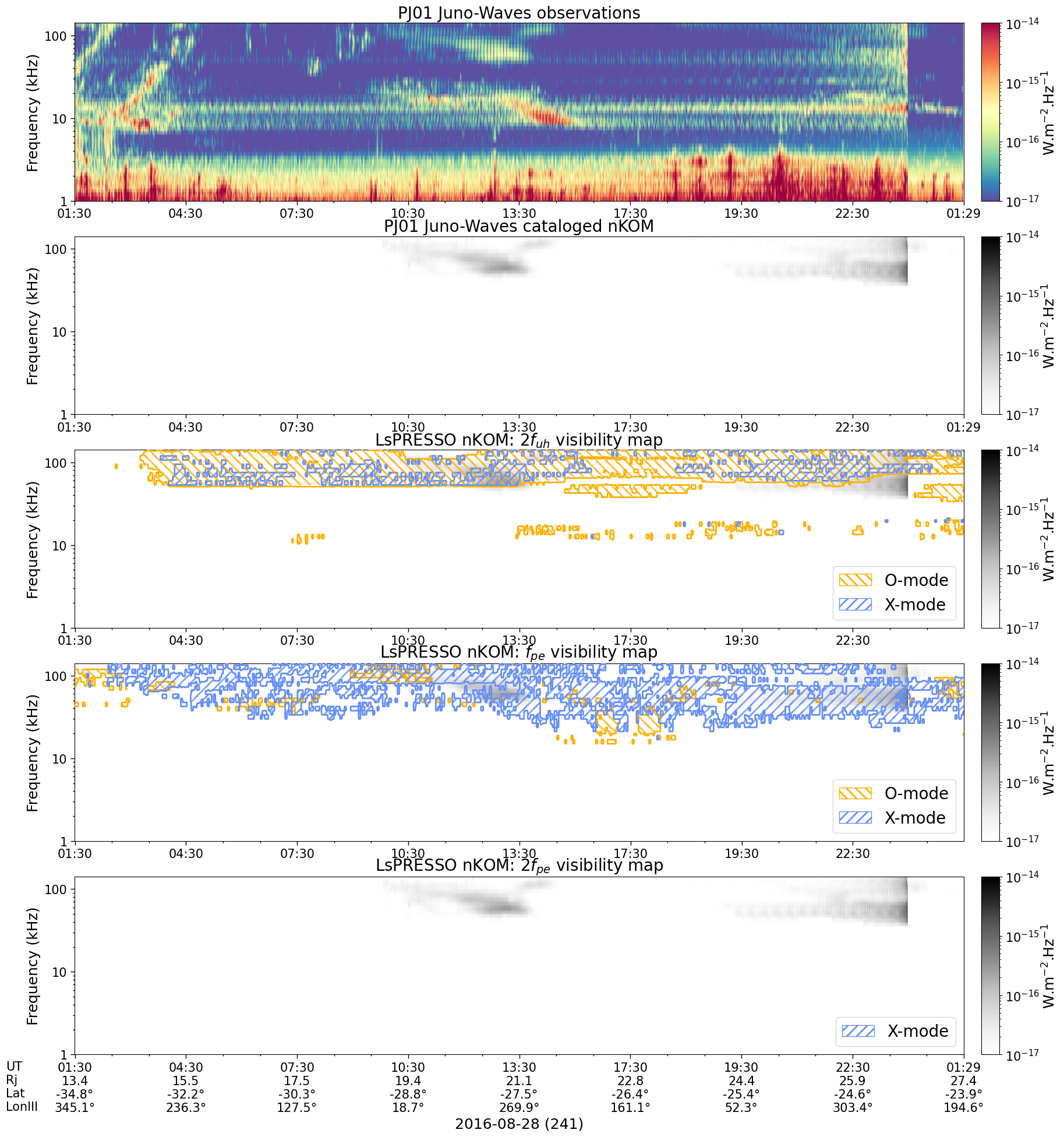}
  \caption{Comparison between nKOM observations by Juno/Waves during PJ01 and the visibility maps predicted by LsPRESSO for the compatible generation scenarios determined from Table \ref{tab:results}. The first panel corresponds to the time-frequency spectrogram of the calibrated Juno/Waves observations. The second panel shows the time-frequency spectrogram of the calibrated nKOM observations by Juno/Waves. The intensity gap between the observations of 28/08/2016 \change{and 28/08/2016}{(right after 00:00) and 29/08/2016} are due to the calibration process. The third, fourth, and fifth panels display the nKOM observations by Juno/Waves, overlaid with the time-frequency visibility maps of nKOM predicted by LsPRESSO for generation scenarios \#2, \#3, and \#4, respectively (see Table \ref{tab:results}). The orange and blue hatched regions correspond to the intercepted beams for O-mode and X-mode cutoffs, respectively. For all panels, the x-axes represent the observation period (UT), radial distance ($R_J$), jovicentric latitude (Lat), and System III longitude (LonIII) of Juno. The y-axes correspond to the Juno/Waves observation frequencies between channels \#16 and \#60 (i.e., from 1~kHz to 141~kHz) on a logarithmic scale.}
\label{fig:visibility_map_PJ01_nKOM}
\end{figure}

\begin{figure}[htbp]
  \centering
  \includegraphics[width=\textwidth]{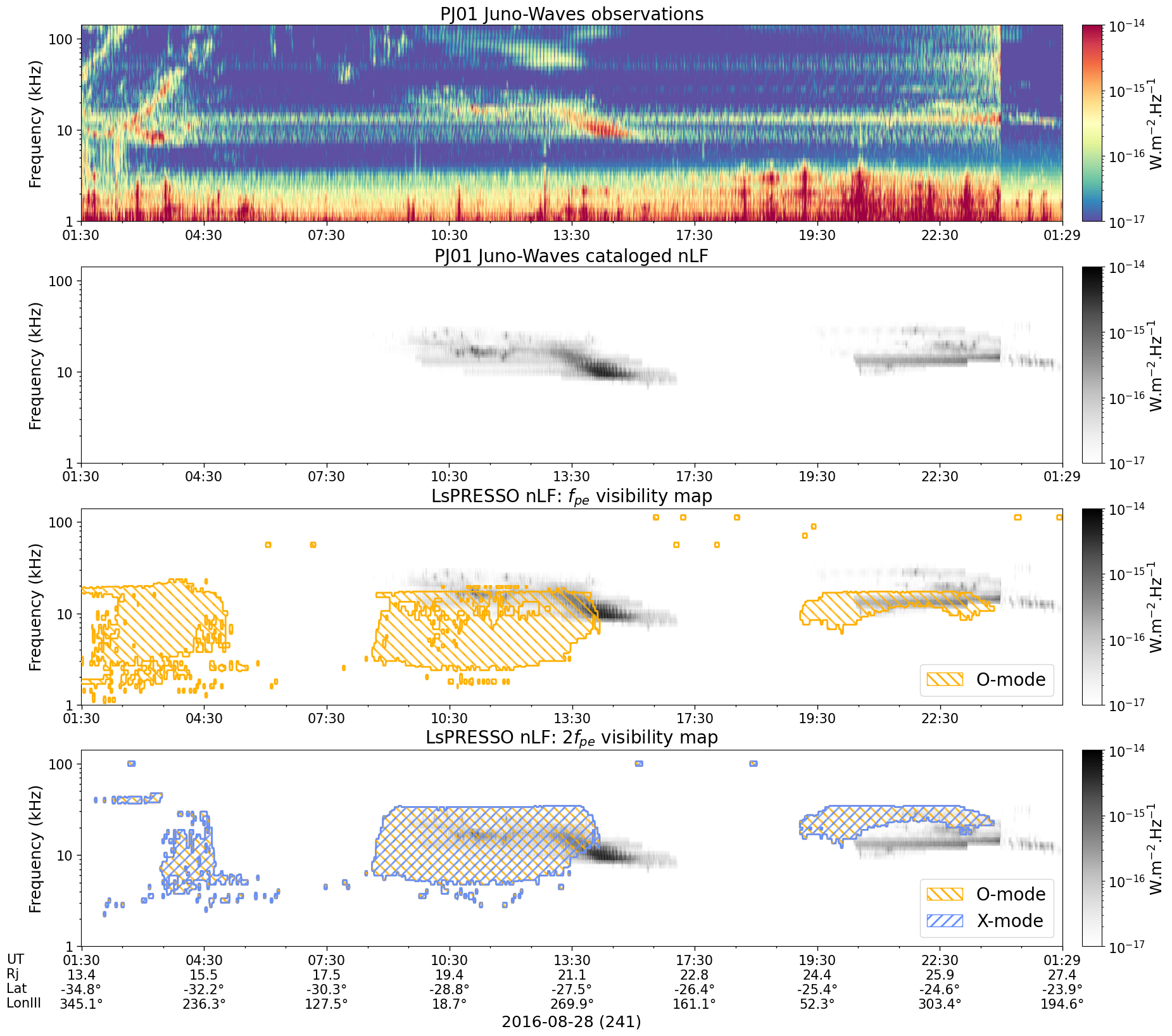}
  \caption{Comparison between nLF observations by Juno/Waves during PJ01 and the visibility maps predicted by LsPRESSO for the compatible generation scenarios determined from Table \ref{tab:results}. The panel definitions are identical to those in Figure \ref{fig:visibility_map_PJ01_nKOM}.}
\label{fig:visibility_map_PJ01_nLF}
\end{figure}

The time-frequency visibility maps simulated with LsPRESSO for each generation scenario can be directly compared with the Juno/Waves spectrograms. Figures \ref{fig:visibility_map_PJ01_nKOM} and \ref{fig:visibility_map_PJ01_nLF} display the Juno/Waves observations on 28/08/2016, the day following PJ01, along with the visibility maps simulated for the generation scenarios that are compatible with the Juno/Waves observations of nKOM and nLF, respectively. In the following paragraphs, the visibility maps predicted by LsPRESSO are compared to the observations made by Juno/Waves for both nKOM and nLF.

%%%%%%%%%%%%%%%%%%%%%%%%%%%%%%%%%%%%%%%%%%%%%%%%%%%%%%%%%%%%%%%%%%%%%%%%%%%%%
%_) nKOM
%_) nKOM Spectrogram
The spectrogram of nKOM observations (2nd panel of Figure \ref{fig:visibility_map_PJ01_nKOM}) shows that two events were catalogued as nKOM at 10h00 and 19h00, occurring between 45--115~kHz.
%%%%%%%%%%%%%%%%%%%%%%%%%%%%%%%%%%%%%%%%%%%%%%%%%%%%%%%%%%%%%%%%%%%%%%%%%%%%%
The visibility maps for the generation scenarios compatible with the nKOM observations only partially predict the visibility of nKOM observed during PJ01. Scenarios \#2, \#3, and \#4 (3rd, 4th, and 5th panels of Figure \ref{fig:visibility_map_PJ01_nKOM}) exhibit similar morphologies, with scenario \#4 appearing to best match the time period and frequency range of the nKOM observations. However, none of the simulated visibility maps reproduce the periodicity of nKOM observations, particularly the extinction between 13h00 and 18h00. Finally, it can be noted that in the low-frequency limit, scenario \#3 with an X-mode cutoff appears to follow the high-frequency boundary of the nLF observations (cf Section \ref{sec:discussion}).
%%%%%%%%%%%%%%%%%%%%%%%%%%%%%%%%%%%%%%%%%%%%%%%%%%%%%%%%%%%%%%%%%%%%%%%%%%%%%

%%%%%%%%%%%%%%%%%%%%%%%%%%%%%%%%%%%%%%%%%%%%%%%%%%%%%%%%%%%%%%%%%%%%%%%%%%%%%
%_) nLF
%_) nLF Spectrogram
The spectrogram of nLF observations (2nd panel of Figure \ref{fig:visibility_map_PJ01_nLF}) shows that two events were catalogued as nLF at 9h00 and 19h00, occurring between 10--25~kHz.
%%%%%%%%%%%%%%%%%%%%%%%%%%%%%%%%%%%%%%%%%%%%%%%%%%%%%%%%%%%%%%%%%%%%%%%%%%%%%
The visibility maps for the generation scenarios compatible with the nLF observations correctly predict the visibility of nLF observed during PJ01. Scenarios \#3 and \#4 (3rd and 4th panels of Figure \ref{fig:visibility_map_PJ01_nLF}) exhibit similar morphologies. Scenario \#4 with an X-mode cutoff appears to best match the frequency range of the nLF observation at 9h00. However, scenarios \#3 and \#4 seem to accurately represent the nLF observation at 19h00: the low-frequency part coincides with scenario \#3, while the high-frequency part aligns well with scenario \#4. Moreover, unlike the visibility maps simulated for the nKOM observations, those simulated for the nLF observations correctly represent the periodicity of nLF observations. Finally, scenario \#4 with an O-mode cutoff also appears to correspond to a signal between 4h30 and 6h00 that was not catalogued.
%%%%%%%%%%%%%%%%%%%%%%%%%%%%%%%%%%%%%%%%%%%%%%%%%%%%%%%%%%%%%%%%%%%%%%%%%%%%%

\section{Conclusion and Discussion}\label{sec:discussion}

We developed the LsPRESSO code, which simulates the visibility of planetary plasma emissions, and used it to study the generation and localization of nKOM and nLF radio sources based on the statistical analysis of Juno/Waves observations. Using the diffusive density model of \citeA{Imai2017} and the VIP4 magnetic field model with current sheet \cite{Connerney1998}, we simulated the observation of Jovian plasma emissions by Juno/Waves exploring four different generation scenarios (see Table \ref{tab:scenario}). Assuming that the activation of radio sources is controlled by the angle $\alpha$ between ${\boldsymbol\nabla} n_e$ and $\mathbf{B}$, as well as the large-scale density gradient magnitude $||\boldsymbol{\nabla} n_e||$ (characterized by its \change{centile}{percentile} $\epsilon$), we compared the latitude-frequency distributions derived from the simulated visibility maps with those observed by Juno/Waves. This allowed us to determine which generation scenarios are capable of reproducing the observations.

%%%%%%%%%%%%%%%%%%%%%%%%%%%%%%%%%%%%%%%%%%%%%%%%%%%%%%%%%%%%%%%%%%%%%%%%%%%%%
Scenario \#1, based on the theory of \citeA{Jones1987}, is incompatible with the Juno/Waves observations of nKOM and nLF. Indeed, the predicted apparent emission directivity following the angle $\beta = \arctan (\sqrt{\omega_{pe}/\omega_{ce}})$ relative to $\mathbf{B}$ does not explain the latitudinal distribution of Juno/Waves observations.
%%%%%%%%%%%%%%%%%%%%%%%%%%%%%%%%%%%%%%%%%%%%%%%%%%%%%%%%%%%%%%%%%%%%%%%%%%%%%

%%%%%%%%%%%%%%%%%%%%%%%%%%%%%%%%%%%%%%%%%%%%%%%%%%%%%%%%%%%%%%%%%%%%%%%%%%%%%
Scenario \#3 in O-mode, which corresponds to emissions generated at $\omega_{pe}$ in the direction $\mathbf{r} \parallel -\boldsymbol{\nabla} n_e$, is compatible with both nKOM and nLF observations by Juno/Waves. This scenario can separately reproduce nKOM observations in ZO (when $\alpha\sim57^{\circ}$ and $\epsilon\sim80\%$) and nLF observations in XO (when $\alpha\sim90^{\circ}$ and $\epsilon<70\%$). Moreover, it is noticeable that the simulated distribution for nLF in XO (see Figure S9.b3 in the Supplementary Information) also presents occurrences at high-latitudes similar to nKOM observations in ZO. Simulations by \citeA{Schleyer2013, Kim2013} have shown that LMC of Langmuir waves to O-mode is efficient for large angles ($\angle(\boldsymbol{\nabla} n_e,\mathbf{B}) > 70^{\circ}$), but also for intermediate angles ($40^{\circ}<\angle(\boldsymbol{\nabla} n_e,\mathbf{B})<60^{\circ}$) when the plasma is moderately magnetized. This suggests that nKOM in ZO and part of nLF in XO are generated at $\omega_{pe}$ in O-mode via LMC mechanisms. Furthermore, the LMC theory of \citeA{Jones1980} predicts efficient conversions when $\boldsymbol{\nabla} n_e \perp \mathbf{B}$, suggesting that near the O-mode cutoff, beam alignment tends to follow the anti-parallel density gradient direction rather than the angle $\beta$.
%%%%%%%%%%%%%%%%%%%%%%%%%%%%%%%%%%%%%%%%%%%%%%%%%%%%%%%%%%%%%%%%%%%%%%%%%%%%%

%%%%%%%%%%%%%%%%%%%%%%%%%%%%%%%%%%%%%%%%%%%%%%%%%%%%%%%%%%%%%%%%%%%%%%%%%%%%%
Scenario \#4, which corresponds to emissions generated at $2\omega_{pe}$ in the anti-parallel density gradient direction, is compatible with the Juno/Waves observations of nLF. This scenario can reproduce nLF observations in ZO in O-mode (when $\alpha<15^{\circ}$ and $\epsilon>90\%$) and nLF observations in XO in both O-mode and X-mode (when $\alpha>75^{\circ}$ and $\epsilon>70\%$). This suggests that nLF in ZO and part of nLF in XO are generated at $2\omega_{pe}$ through NLMC mechanisms. Since emissions at $2\omega_{pe}$ can occur far from the O-mode and X-mode cutoffs, it is difficult to explain without ray-tracing models why the beam directivity is $\mathbf{r} \parallel -\boldsymbol{\nabla} n_e$. However, as these emissions are generated when $\boldsymbol{\nabla} n_e \parallel |\mathbf{B}|$ (for nLF in ZO) and $\boldsymbol{\nabla} n_e \perp |\mathbf{B}|$ (for nLF in XO), their directivities may correspond to $\mathbf{r} \parallel |\mathbf{B}|$ and $\mathbf{r} \perp |\mathbf{B}|$, respectively.
%%%%%%%%%%%%%%%%%%%%%%%%%%%%%%%%%%%%%%%%%%%%%%%%%%%%%%%%%%%%%%%%%%%%%%%%%%%%%

%%%%%%%%%%%%%%%%%%%%%%%%%%%%%%%%%%%%%%%%%%%%%%%%%%%%%%%%%%%%%%%%%%%%%%%%%%%%%
Scenario \#2, based on the theory of \citeA{Fung&Papadopoulos1987}, along with scenarios \#3 and \#4, are compatible with some of the Juno/Waves observations of nKOM. Scenario \#2 in both O-mode and X-mode, and scenarios \#3 and \#4 in X-mode, can reproduce nKOM observations in XO (when $\alpha>30^{\circ}$). The simulated distributions in X-mode for scenarios \#3 and \#4 (see Figures S4.b3 and b4 in the Supplementary Information) are qualitatively more similar than those simulated under scenario \#2 (see Figures S3.b2 and S4.b2 in the Supplementary Information). They are also able to reproduce the absence of occurrences around $\sim100$~kHz. However, the results remain quantitatively ambiguous and do not allow for a conclusive determination of the generation and directivity of nKOM in XO.
%%%%%%%%%%%%%%%%%%%%%%%%%%%%%%%%%%%%%%%%%%%%%%%%%%%%%%%%%%%%%%%%%%%%%%%%%%%%%

%%%%%%%%%%%%%%%%%%%%%%%%%%%%%%%%%%%%%%%%%%%%%%%%%%%%%%%%%%%%%%%%%%%%%%%%%%%%%
Finally, for the generation scenarios compatible with the Juno/Waves observations of nKOM and nLF, the radio sources and time-frequency visibility maps predicted by LsPRESSO were compared with goniometry measurements and spectrograms obtained at PJ01 by \citeA{Imai2017}. For nKOM, the equatorial distribution of radio sources does not predict the azimuthal asymmetry observed by Juno/Waves, and the visibility maps do not reproduce the $\sim10.5$-hour periodicity. However, both of these properties are correctly reproduced in the case of nLF.
%%%%%%%%%%%%%%%%%%%%%%%%%%%%%%%%%%%%%%%%%%%%%%%%%%%%%%%%%%%%%%%%%%%%%%%%%%%%%

%%%%%%%%%%%%%%%%%%%%%%%%%%%%%%%%%%%%%%%%%%%%%%%%%%%%%%%%%%%%%%%%%%%%%%%%%%%%%
This difference may be explained by the intermittency of nKOM. Observations by Galileo have shown that the activation of nKOM radio sources is constrained by energetic events in Jupiter’s magnetosphere \cite{Louarn1998, Louarn2000, Louarn2001}. However, in this study, we assumed that plasma radio source activation is permanent. This implies that in the simulated visibility maps, only a fraction of the considered sources were actually active. Finally, although the comparison between LsPRESSO simulations and \citeA{Imai2017} estimations gives new insights on the radio sources locations, the assumptions on the goniometry method have limitations, especially for the meridian case, as they assumed generation at $\omega_{pe}$ and deduced the projection employing the same time-averaged diffusive density model as the one used in this study \cite{Imai2016}.
%%%%%%%%%%%%%%%%%%%%%%%%%%%%%%%%%%%%%%%%%%%%%%%%%%%%%%%%%%%%%%%%%%%%%%%%%%%%%

%%%%%%%%%%%%%%%%%%%%%%%%%%%%%%%%%%%%%%%%%%%%%%%%%%%%%%%%%%%%%%%%%%%%%%%%%%%%%
On the other hand, since this issue is not observed with nLF, it suggests that the radio sources detected during the southern plasma disk crossing are more persistent. This implies that nLF could be less intermittent than nKOM, and its generation may not be primarily controlled by magnetospheric disturbances.
%%%%%%%%%%%%%%%%%%%%%%%%%%%%%%%%%%%%%%%%%%%%%%%%%%%%%%%%%%%%%%%%%%%%%%%%%%%%%

These nLF observations correspond to plasma emissions generated at $\omega_{pe}$ and $2\omega_{pe}$. They originate from the same radio sources at $>10~R_J$ where ${\boldsymbol\nabla} n_e \perp \mathbf{B}$, indicating the coexistence of LMC and NLMC mechanisms. In \citeA{Louis2021}, the latitude-frequency distribution of nLF intensity suggests that nLF events are more intense at mid-latitudes when $<15$~kHz \cite<see Figure 4.b of>[]{Louis2021}. The predicted visibility maps for PJ01 (see Figure \ref{fig:visibility_map_PJ01_nLF}) also show that at 16h00, the observed event matching $\omega_{pe}$ is more intense than the one matching $2\omega_{pe}$. This could suggest that the linear conversion process dominates over nonlinear processes, indicating that the radio sources exhibit density fluctuations similar to those simulated by \citeA{Krafft&Savoini2021, Krafft&Volokitin2024, Krafft2024}.
\appendix

%%%%%%%%%%%%%%%%%%%%%%%%%%%%%%%%%%%%%%%%%%%%%%%%%%%%%%%%%%%%%%%%%%%%%%%%%%%%%
\section{Propagation Mode Identification using the JADE and MAG instruments}\label{appendix:dispersion}
%%%%%%%%%%%%%%%%%%%%%%%%%%%%%%%%%%%%%%%%%%%%%%%%%%%%%%%%%%%%%%%%%%%%%%%%%%%%%

%%%%%%%%%%%%%%%%%%%%%%%%%%%%%%%%%%%%%%%%%%%%%%%%%%%%%%%%%%%%%%%%%%%%%%%%%%%%%
%_On suppose un plasma froid [dispersion ANNEXE]
%_Relation de dispersion: chaque mode est délimité en fréquences en fonction de (fpe, fce)
Assuming a cold collisionless magnetized plasma, we can derive the frequency ranges of the normal modes of the plasma as a function of $\omega_{pe}$ and $\omega_{ce}$ using the \textit{Altar-Appleton-Hartree} relation \cite{Appleton1932}:
\begin{equation}\label{eq:dispersion}
    \epsilon(\omega) = 1-\frac {2X(1-X)}{2(1-X) - Y^2\sin^2\theta \pm \sqrt{(Y^2 \sin^2\theta)^2 + 4Y^2 (1 - X^2)\cos^2\theta}}
\end{equation}
with $X = \omega_{pe}^2/\omega^2$ and $Y = \omega_{ce}/\omega$.

\begin{figure}[!ht]
\centering
\noindent\includegraphics[width=\textwidth]{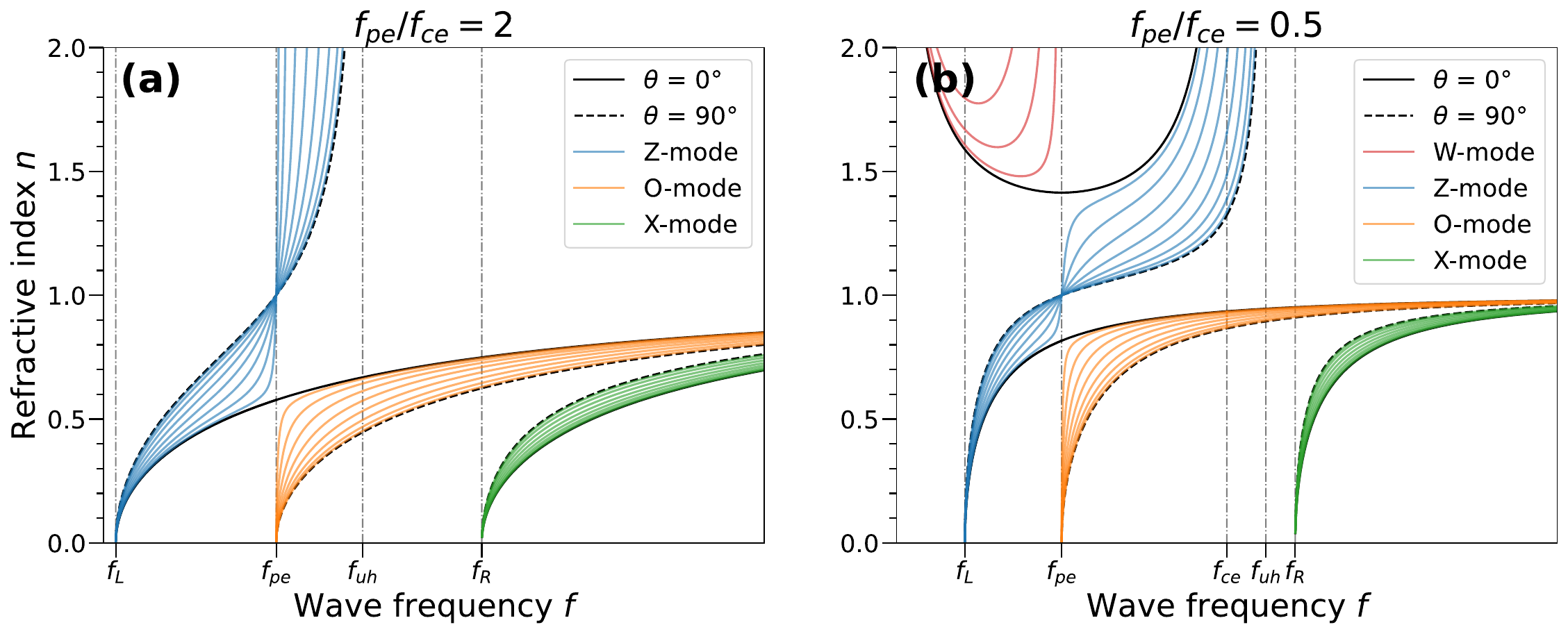}
\caption{Dispersion curves of a cold, homogeneous, collisionless plasma immersed in a uniform magnetic field. Panels (a) and (b) correspond to the solutions of the dispersion relation defined by Equation (\ref{eq:dispersion}) for different values of $\theta$ when $\omega_{pe}/\omega_{ce}=2$ and $\omega_{pe}/\omega_{ce}=1/2$ with $\omega=2\pi f$. The red, blue, orange, and green curves correspond to the W, Z, O, and X modes, respectively, for $\theta \in]0^{\circ}, 90^{\circ}[$. The solid and dashed black curves correspond to $\theta=0^{\circ}$ and $\theta=90^{\circ}$, respectively.}
\label{appendix:fig:dispersion_curve}
\end{figure}

\begin{figure}[!ht]
\centering
\noindent\includegraphics[width=\textwidth]{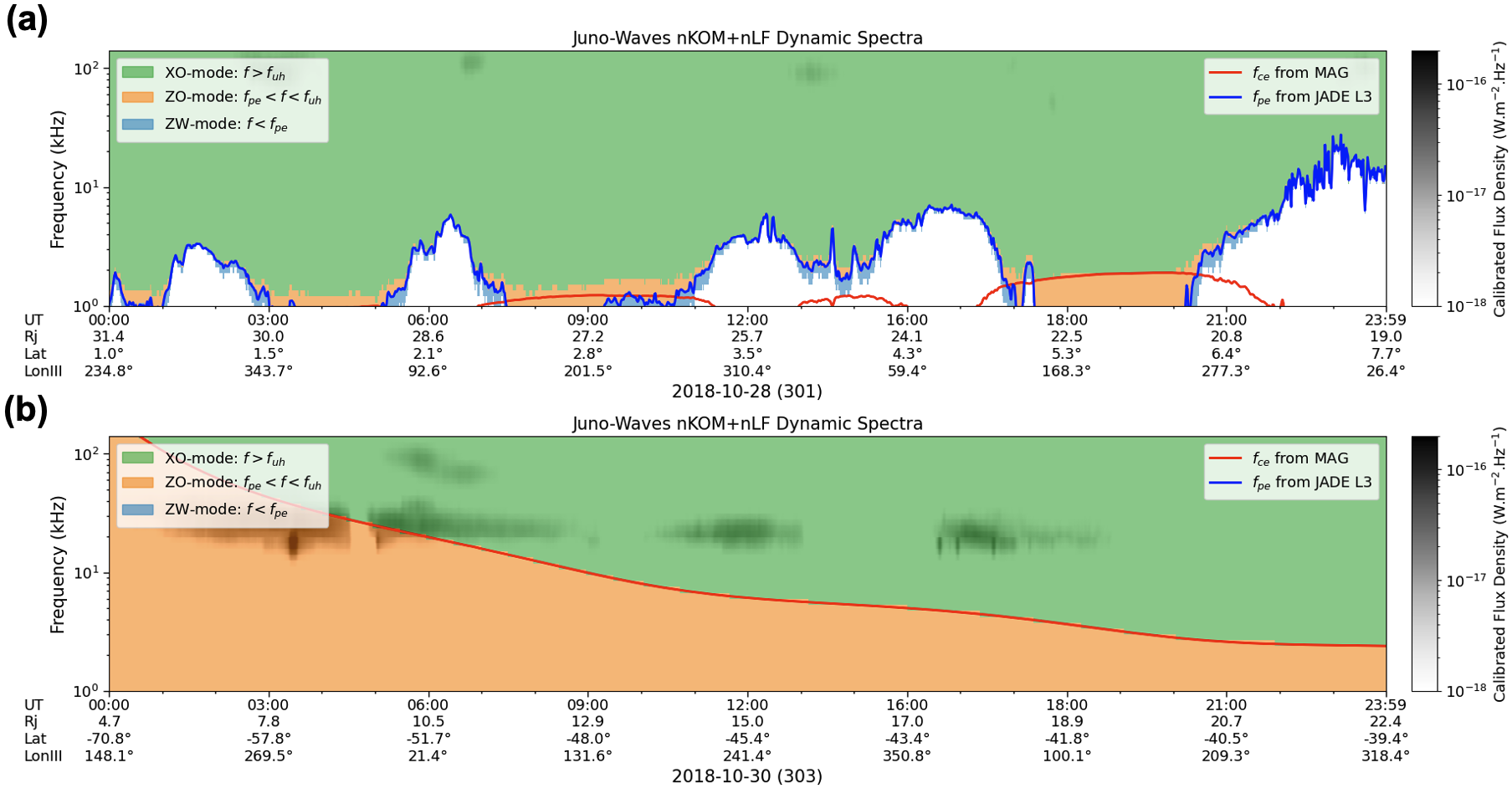}
\caption{24-hour spectrograms of the calibrated nKOM and nLF observations by Waves compared to the ZW, XO, and ZO spectral ranges in the plasma disk (a) on the day before (28/10/2018) and (b) on the day after (30/10/2018) PJ16. The x- and y-axes are identical to those in Figure \ref{fig:spdyn_juno}. The blue, orange, and green areas correspond to the ZW, ZO, and XO frequency ranges, respectively, estimated from JADE and MAG data. The red curve represents the electron cyclotron frequency $f_{ce}=\omega_{ce}/2\pi$ derived from MAG data. The blue curve represents the plasma frequency $f_{pe}=\omega_{pe}/2\pi$ derived from JADE data.}
\label{appendix:fig:lat_vs_freq_dist:modes}
\end{figure}

Figure \ref{appendix:fig:dispersion_curve} illustrates this for two different ratio of $\omega_{pe}$ over $\omega_{ce}$ : (a) for $\omega_{pe}/\omega_{ce}=2$, (b) $\omega_{pe}/\omega_{ce}=1/2$. Figure \ref{appendix:fig:lat_vs_freq_dist:modes} shows the 24-hour spectrograms of the calibrated nKOM and nLF observations (a) on the day before (28/08/2018) and (b) on the day after (30/08/2018) PJ16, overlaid with the frequency ranges in ZW, XO, and ZO (see their definitions Section \ref{sec:juno_data:propagation_modes}). Panel (a) indicates that during the northern crossing of the plasma disk, all nKOM events are observed in XO. Panel (b) shows that during the southern crossing of the plasma disk, the nLF event at 03h00 is observed in ZO, whereas all nKOM and nLF events after 06h00 are observed in XO. However, the nLF events at 03h00 and 06h00 appear to be connected and were individually contoured due to the presence of QP at 05h00 (see Figure \ref{fig:spdyn_juno}). In this case, as the nLF event is at the intersection between ZO and XO, this suggests that it propagates in O-mode.

%%%%%%%%%%%%%%%%%%%%%%%%%%%%%%%%%%%%%%%%%%%%%%%%%%%%%%%%%%%%%%%%
%
% Optional Glossary, Notation or Acronym section goes here:
%
%%%%%%%%%%%%%%
% Glossary is only allowed in Reviews of Geophysics
%  \begin{glossary}
%  \term{Term}
%   Term Definition here
%  \term{Term}
%   Term Definition here
%  \term{Term}
%   Term Definition here
%  \end{glossary}

%
%%%%%%%%%%%%%%
% Acronyms
%   \begin{acronyms}
%   \acro{Acronym}
%   Definition here
%   \acro{EMOS}
%   Ensemble model output statistics
%   \acro{ECMWF}
%   Centre for Medium-Range Weather Forecasts
%   \end{acronyms}

%
%%%%%%%%%%%%%%
% Notation
%   \begin{notation}
%   \notation{$a+b$} Notation Definition here
%   \notation{$e=mc^2$}
%   Equation in German-born physicist Albert Einstein's theory of special
%  relativity that showed that the increased relativistic mass ($m$) of a
%  body comes from the energy of motion of the body—that is, its kinetic
%  energy ($E$)—divided by the speed of light squared ($c^2$).
%   \end{notation}

%%%%%%%%%%%%%%%%%%%%%%%%%%%%%%%%%%%%%%%%%%%%%%%

\section*{Data Availability Statement}

The Juno/Waves calibrated data used in the manuscript is available at \url{https://doi.org/10.25935/6jg4-mk86} \cite{Louis2021_data}. The Juno/Waves catalog used in the manuscript is available at \url{https://doi.org/10.25935/nhb2-wy29} \cite{Louis2021_catalog}.

\acknowledgments
A. Boudouma's work at LIRA was supported by the CNES (Centre National d'Etudes Spatiale). A. Boudouma's work at the Department of Space Physics of the Institute of Atmospheric Physics is supported by the Lumina Quaeruntur fellowship (LQ100422501) of the Czech Academy of Sciences.

%% ------------------------------------------------------------------------ %%
%% References and Citations

%%%%%%%%%%%%%%%%%%%%%%%%%%%%%%%%%%%%%%%%%%%%%%%
%
\bibliography{agusample}
%
% don't specify bibliographystyle

% In the References section, cite the data/software described in the Availability Statement (this includes primary and processed data used for your research). For details on data/software citation as well as examples, see the Data & Software Citation section of the Data & Software for Authors guidance
% https://www.agu.org/Publish-with-AGU/Publish/Author-Resources/Data-and-Software-for-Authors#citation

%%%%%%%%%%%%%%%%%%%%%%%%%%%%%%%%%%%%%%%%%%%%%%%

%\bibliography{enter your bibtex bibliography filename here}

%Reference citation instructions and examples:
%
% Please use ONLY \cite and \citeA for reference citations.
% \cite for parenthetical references
% ...as shown in recent studies (Simpson et al., 2019)
% \citeA for in-text citations
% ...Simpson et al. (2019) have shown...
%
%
% ...as shown by \citeA{Kendall}.
% ...as shown by \citeA{lewin76}, \citeA{carson86}, \citeA{bartoldy02}, and \citeA{rinaldi03}.
%...has been shown \cite{jskilbye}.
%...has been shown \cite{lewin76,carson86,bartoldy02,rinaldi03}.
%... \cite <i.e.>[]{lewin76,carson86,bartoldy02,rinaldi03}.
%...has been shown by \cite <e.g.,>[and others]{lewin76}.
%
% apacite uses < > for prenotes and [ ] for postnotes
% DO NOT use other cite commands (e.g., \citet, \citep, \citeyear, \citealp, etc.).
% \nocite is okay to use to add references from your Supporting Information
%

\end{document}